%% file: main.tex
\RequirePackage{luatex85}
\documentclass[10pt,journal,compsoc]{IEEEtran}

\ifCLASSOPTIONcompsoc{}
  \usepackage[nocompress]{cite}
\else
  \usepackage{cite}
\fi

\usepackage[pdftex]{graphicx}
\usepackage{mathtools}
\usepackage{amsmath,amsfonts}
\usepackage[inline]{enumitem}
\usepackage{algorithm}
\usepackage{algorithmicx}
\usepackage{algpseudocode}
\usepackage{algxpar}
\usepackage{stfloats}
\usepackage{array}
\usepackage{url}
\usepackage{xcolor}
\usepackage{enumitem}
\usepackage{rotating}
\newcommand{\R}{\mathbb{R}}

\title{ Grounds for Suspicion: Physics-based Early Warnings for Stealthy Attacks on Industrial Control Systems }

\author{Mazen~Azzam,
        Liliana~Pasquale,
        Gregory~Provan,
        and~Bashar~Nuseibeh
        \IEEEcompsocitemizethanks{\IEEEcompsocthanksitem M. Azzam (mazen.azzam@ul.ie) and B. Nuseibeh (bashar.nuseibeh@ul.ie are with Lero, the Irish Software Research Centre, University of Limerick, Limerick, Ireland.\protect\\
        \IEEEcompsocthanksitem L. Pasquale (liliana.pasquale@ucd.ie is with Lero, University College Dublin, Dublin, Ireland.\protect\\
        \IEEEcompsocthanksitem G. Provan (g.provan@cs.ucc.ie) is with Lero, University College Cork, Cork, Ireland.}}

\begin{document}
%%%%%%%%%%%%%%%%%%%%%%%%%%%%%%%%%%%%%%%%%%%%%%%%%%%%%%%%%%%%%%%%%%%%%%%%%%%%%%%%
\IEEEtitleabstractindextext{
\begin{abstract}

\textit{Stealthy attacks} on Industrial Control Systems can cause significant damage while evading detection. In this paper, instead of focusing on the detection of stealthy attacks, we aim to provide early warnings to operators, in order to avoid physical damage and preserve in advance data that may serve as an evidence during an investigation. We propose a framework to provide \textit{grounds for suspicion}, i.e. preliminary indicators reflecting the likelihood of success of a stealthy attack. We propose two grounds for suspicion based on the behaviour of the physical process: (i) \textit{feasibility} of a stealthy attack, and (ii) \textit{proximity} to unsafe operating regions. We propose a metric to measure grounds for suspicion in real-time and provide soundness principles to ensure that such a metric is consistent with the grounds for suspicion. We apply our framework to Linear Time-Invariant (LTI) systems and formulate the suspicion metric computation as a real-time reachability problem.  We validate our framework on a case study involving the benchmark Tennessee-Eastman process. We show through numerical simulation that we can provide early warnings well before a potential stealthy attack can cause damage, while  incurring minimal load on the network. Finally, we apply our framework on a use case to illustrate its usefulness in supporting early evidence collection.

\end{abstract}

}

%%%%%%%%%%%%%%%%%%%%%%%%%%%%%%%%%%%%%%%%%%%%%%%%%%%%%%%%%%%%%%%%%%%%%%%%%%%%%%%%
\maketitle

\input{section1.Introduction.tex}
\input{section2.Related_Work.tex}
\input{section3.Motivating_Example}
\input{section4.Contribution.tex}
\input{section5.Case_Study.tex}
\input{section6.Instantiating.tex}
\input{section7.Evaluation.tex}
\input{section8.Conclusion.tex}

\section*{Acknowledgements}
This work was supported by Science Foundation Ireland (SFI).
\bibliographystyle{IEEEtran}
\bibliography{IEEEabrv,libs/lib_cpssecurity.bib,libs/lib_dynamicalsystems.bib,libs/lib_forensics.bib,libs/lib_suspicion.bib}

\end{document}

%% file: section1.Introduction.tex
\IEEEraisesectionheading{\section{Introduction}\label{section:introduction}}
%\begingroup
\IEEEPARstart{C}{yber-Physical Systems} (CPS) augment physical systems with enhanced capabilities, such as real-time monitoring and dynamic control~\cite{Alur_Principles_2015}. Industrial Control Systems (ICS) are considered a subclass of CPS, where software controls safety-critical industrial processes. Attacks against ICS can have disruptive consequences to users and physical assets, as shown by the German steel mill attack in 2014~\cite{Lee_SANS_2014} and the attack against the Ukrainian power grid in 2015~\cite{Lee_Analysis_2016} --- among others.
\par
Anomaly-based Intrusion Detection Systems (IDS) can usually detect attacks affecting the physical process in an ICS, by monitoring deviations from the normal system behaviour (anomalies)~\cite{Giraldo_Survey_2018}. However, skilled attackers can take advantage of the noise in the system and the thresholds used by the anomaly detectors, to cause damage to the ICS before an alarm is raised~\cite{Bai_Security_2015,Pasqualetti_Attack_2013}. Such attacks which evade detection are also known as \emph{stealthy attacks}. Early Warning Systems (EWS)~\cite{Apel_Towards_2009,Kalutarage_Early_2015} traditionally monitor the occurrence of suspicious and seemingly benign network events (often called weak evidence). Differently from IDS, EWS generate predictions and advice on unfamiliar situations before a potential attack can cause harm~\cite{Kalutarage_Early_2015}. EWS may not reveal attacks on their own, but can guide the selection of appropriate measures to detect potential intrusions, and as such complement existing IDS as a security solution~\cite{Ramaki_Survey_2016}. In this paper, instead of forcing the detection of stealthy attacks, we aim to raise early warnings when there is sufficient evidence that a potential stealthy attack can cause damage.
\par
Our main contribution is a framework to generate early warnings in ICS based on preliminary indicators of a stealthy attack, referred to as \emph{grounds for suspicion}. This framework can be used within a larger EWS which considers indicators from other sources. The success of a stealthy attack depends on the laws of physics underlying the behaviour of the ICS and the anomaly detector. Thus, we define two grounds for suspicion based on the physical state of the system: (i) \emph{Feasibility} of a stealthy attack indicates whether the ICS can be taken to an unsafe operating region, while avoiding detection by the IDS.\@ (ii) \emph{Proximity} represents the vicinity of the system to the unsafe operating region. %Intuitively, a higher proximity can indicate an increased chance of success for an attacker to stealthily damage the system. 
To monitor the grounds for suspicion, we propose a \textit{suspicion metric} based on a mathematical model of the system and a notion of reachability. We also provide soundness principles to ensure that a metric is consistent with the measured grounds for suspicion.
\par
To assess feasibility of our framework, we study its applicability to Linear Time-Invariant (LTI) systems, a standard physical modelling framework commonly used in process control. We adapt existing reachability analysis tools~\cite{Murguia_reachable_2018} to compute the suspicion metric. We alleviate the computational cost of performing real-time reachability analysis by computing symbolic reachable sets of system states offline~\cite{Chen_Model_2017}. We then instantiate these sets online given a prediction of the physical state variables for a certain number of time steps into the future. We leverage ellipsoidal techniques~\cite{kurzhanski_ellipsoidal_2000,kurzhanskiy_ellipsoidal_2006,boyd_convex_2004} to perform efficient safety checks online and compute the suspicion metric. In a previous work~\cite{Azzam_Efficient_2021}, we focused on performing safety checking for LTI systems under stealthy attacks in real-time. In the present work, we extend these results into an algorithm that efficiently computes the suspicion metric online, and we design suitable thresholds for warnings of different criticality in a way that fits the soundness principles of proposed framework.
\par
We validate our framework's application to LTI systems using a testbed involving a networked version of the benchmark Tennessee-Eastman Process (TEP). We use numerical simulations to showcase that our framework can generate early warnings before a stealthy attack can cause damage. Although we perform this study using a particular type of stealthy attack --- false data injection on sensors --- our framework is generalizable to other types of stealthy attacks. Furthermore, we demonstrate that our framework scales well with the number of safety constraints and incurs minimal load on the network. Finally, we apply our framework on a use case inspired by the TEP benchmark, to showcase its usefulness in supporting early evidence collection, especially from low-level control devices. However, these benefits come with a cost associated with the human effort required to instantiate the framework. 
\par
The rest of the paper is organised as follows. Section~\ref{section:related_work} provides a brief overview of related work. Section~\ref{section:motivation} illustrates a motivating example, while Section~\ref{section:contribution} describes the main contribution of the paper, which the framework for physics-based early warnings. We begin our case study in Section~\ref{section:system} where we introduce the TE benchmark. In Section~\ref{section:instance} we explain how we instantiated our framework to the benchmark. Section~\ref{section:usecase} presents our evaluation results, and Section~\ref{section:concl} concludes the paper.
%\endgroup

%% file: section2.Related_Work.tex
\section{Related Work}\label{section:related_work}
%\begingroup
In this section, we provide some background on existing attack
detection techniques in ICS.\@ We also clarify the positioning of the
paper with respect to existing work on early warning systems and
attack impact assessment in CPS/ICS.\@

\subsection{Attack Detection in ICS}
%\begingroup
Several network-based intrusion detection systems for CPS, and particularly for ICS, have been suggested in previous work. Some of
them~\cite{Genge_connection_2014,Cheminod_Detection_2017} are
knowledge-based and look for features in the network traffic that are
consistent with a known threat
model. Others~\cite{Sayegh_SCADA_2014,Mercaldo_Real_2019} are
anomaly-based and look for features that suggest a deviation from the
expected behaviour. Furthermore, physics-based
methods~\cite{Giraldo_Survey_2018} consider the effect of attacks on
the controlled physical process, and look for deviations from expected
physical sensor measurements, given by a mathematical model of the
system.
\par
However, with enough knowledge about the system, an offender can
launch stealthy attacks. These attacks are usually performed by
introducing fake sensor measurements or actuation signals in the
control loop. In this way, the anomaly detector will not be able to
detect a deviation of the system from the normal
behaviour. Stealthiness of such attacks can be ensured by mimicking
the noise native to the system~\cite{Bai_Security_2015} or exploiting
some control-theoretic
properties~\cite{Pasqualetti_Attack_2013,Teixeira_Revealing_2012}
(e.g., zero dynamics). ``Active'' detection methods have been proposed
to detect stealthy attacks. These methods involve the introduction of
a probing signal (watermark) to reveal fake sensor measurement or
actuation
signals~\cite{Griffioen_Tutorial_2019,Weerakkody_Active_2017}. Active
methods can also be designed for attacks exploiting specific
control-theoretic properties. In this case, detection relies on
modifying the system by, for example, including additional sensor
measurements~\cite{Teixeira_Revealing_2012} or modulating actuation
signals~\cite{Hoehn_Detection_2016}. These modifications can remove the
control-theoretic properties exploited by the attack.
\par
However, active attack detection techniques may bring trade-offs that can
jeopardise their effectiveness. In the case of physical
watermarking, the trade-off between the watermark's robustness and the
control performance may not be acceptable, especially in
safety-critical systems. Also, modifying the structure of the
system (e.g., by adding sensor measurements) is often hard and expensive.
\par
%\endgroup

\subsection{Early Warning Systems}
%\begingroup
EWS combine preventive measures, such as risk and vulnerability assessment, with IDS, to provide a clearer ``picture'' of the security situation and send warnings about potential network intrusions~\cite{Apel_Towards_2009}. Differently from Intrusion/Anomaly Detection Systems (IDS/ADS), EWS collect, accumulate, and combine data from heterogeneous sources in order to raise warnings of potential intrusions. They are regarded as a complementary solution to existing IDS/ADS, where their main benefit lies in providing predictions of potential harm in unfamiliar situations, typically with zero-day attacks~\cite{Ramaki_Survey_2016}.
\par
Recently, a growing body of work has considered this approach in traditional IT systems to tackle slow and stealthy attacks~\cite{Ramaki_Survey_2016}. Apel et al.~\cite{Apel_Early_2010} proposed an EWS that relies on intelligence sharing between different organisations to counter
advanced coordinated attacks at their early stages. Kalutarage et
al.~\cite{Kalutarage_How_2013,Kalutarage_Towards_2015} proposed a Bayesian approach, which
accumulates evidence over long periods of time to counter network
attacks in their reconnaissance phase. The reader is referred to the work by Ramaki et al.~\cite{Ramaki_Survey_2016} for a comprehensive survey of existing work on EWS and a more detailed comparison of EWS and IDS/ADS.
\par
To the best of our
knowledge, a very limited number of approaches apply EWS to ICS. One exception is the recent
work by Abbaszadeh et al.~\cite{Abbaszadeh_Forecasting_2018}, which
generates early warnings based on potential anomalies predicted by
learning time series behaviour. Instead of relying on training machine learning models, we use in our work ideas from reachability analysis based on a standard identified model of the system.
\par
Other related work in the context of ICS/CPS has proposed online monitoring techniques based on a notion of proximity to a predefined set of unsafe or critical states~\cite{Etigowni_Crystal_2018,Carcano_multidimensional_2011,Coletta_Predictive_2018,Castellanos_modular_2019}. For example, Carcano and Coletta~\cite{Carcano_multidimensional_2011,Coletta_Predictive_2018} proposed the use of distance metrics such as Euclidean distance to a set of unsafe states represented as boolean expressions over state variables. Castellanos and Zhou~\cite{Castellanos_modular_2019} further extended this notion by computing an  approximate ``time-to-critical'' states metric, based on Euclidean distance and the rate of change of the physical states. Similarly to these approaches, we compute proximity from the current state of the system which is estimated based on received sensor values. However, this estimated state may not be representative of the state of the system if the latter is under a stealthy attack. Thus, our notion of proximity bounds --- with a certain confidence level --- the actual state of the system. To generate early warnings before a stealthy attack can cause damage, we also predict the state of the system for a certain number of time steps in the future. This prediction is similar to the work by Etigowni et al.~\cite{Etigowni_Crystal_2018} and relies on an approximated model of the system and the monitoring of controller states.

%The main difference with the present work is that we consider the effect of a stealthy attack on the proximity of the system to unsafe states. The common aspect with the listed works is that proximity is computed from the current state of the system which is estimated based on received sensor values. However, this estimated state may not be representative of the state of the system if the latter is under a stealthy attack. In our instantiation of the proposed framework for an ICS, instead of relying on raw sensor values and resulting estimates, the notion of proximity bound --- with a certain confidence level --- the actual state of the system. 
\par
Bradford et al.~\cite{Bradford_Towards_2004} proposed the idea of a
tiered approach for EWS.\@ First, they profile agents in a system by
accumulating preliminary data, then they perform more detailed
investigations and intensive data collection if some pre-defined
thresholds are crossed. Chivers et
al.~\cite{Chivers_Accumulating_2009} implemented a layered approach for
insider attacks in networked systems, while Kalutarage et
al.~\cite{Kalutarage_Sensing_2012} did it for cyber conflict
attribution. Differently from this work,  we focus on
generating early warnings about stealthy attacks in
ICS.\@ The novelty of our approach lies on the measurement of grounds for
suspicion based on the physical state of the ICS.\@

\subsection{Attack Impact Assessment in CPS/ICS}
%\begingroup
Our work builds on recent research assessing the physical impact of
stealthy attacks on CPS and ICS.\@ In particular, we adapt techniques
based on reachability
analysis~\cite{Murguia_reachable_2018,Murguia_Security_2018,Kwon_Reachability_2018}
to provide measures of the proposed grounds for suspicion. While in
previous work such techniques were mainly employed to assess the
security of a system and perform offline risk analysis, here we adapt
them for online monitoring.
\par
Existing risk assessment approaches in the context of ICS security are based on the assumption that
the system will have steady states when subjected to an
attack. However, this assumption can fail when long
transients are experienced in operating conditions. This is
true in the process industry, where changes in operating
conditions are frequent due to external disturbances and real-time
optimisation requirements~\cite{Sequeira_Real_2002}.
\par
In our instantiation of the framework to LTI systems, our approach is similar to that of Kwon et al.~\cite{Kwon_Reachability_2018}. However, the main difference is that Kwon’s algorithm is designed specifically to cater for Unmanned Areal Vehicles (UAV) applications, where safety constraints are time-varying and have a different mathematical expression to constraints typically found in process control applications that we consider. Furthermore, Kwon’s algorithm requires updating the reachable ellipsoid at each time step through a recursive structure, which may be resource-intensive as it is not clear it can scale well with a large number of state variables typically found in process control applications.
\par
Finally, several approaches exist to assess the impact of stealthy
attack strategies in CPS/ICS.\@ For example, on the one hand, Milosevic et
al.~\cite{Milosevic_Quantifying_2018} propose the use of the infinity
norm of critical states after a certain time period and present a framework for security measure allocation offline given attack complexity and impact measures. Urbina et al.~\cite{Urbina_Limiting_2016}, on the other hand, consider the rate of change of the physical variables under a stealthy attack as a measure of impact.  In this paper, we do not focus our attention on computing the impact of a stealthy attack. Instead, we generate early warnings in real-time depending on the likelihood of a stealthy attack to be successful.
%\endgroup

%\endgroup
%%% Local Variables:
%%% mode: latex
%%% TeX-master: "main"
%%% End:

%% file: section3.Motivating_Example.tex
\section{Motivating Example}\label{section:motivation}

\begin{figure}[!t]
    \centering
    \includegraphics[scale = 0.4]{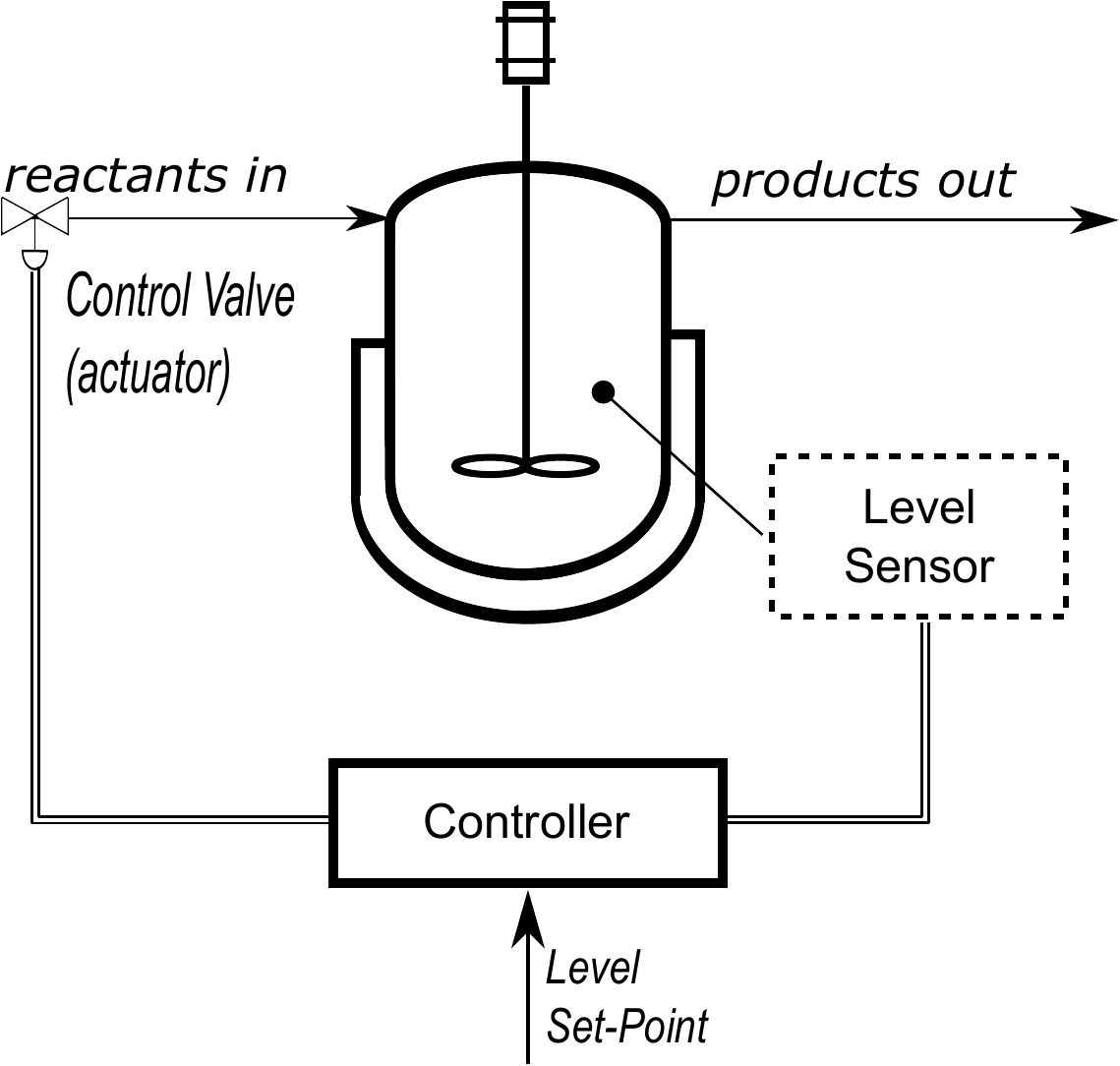}
    \caption{Reactor schematic.}~\label{fig:system.reactor}
\end{figure}
Consider a chemical reactor equipped with a controller that keeps the
level of liquids in the reactor at a desired set-point
(Figure~\ref{fig:system.reactor}). An attacker with access to channels
communicating level values to the controller, wishes to cause
physical damage to the system. To maximise chances of success and
avoid detection, the
attacker uses his/her knowledge of the physical behaviour of the
system (obtained through reconnaissance activity) and its anomaly detector.
\par
\begin{figure}[!t]
    \centering
    \includegraphics[width=\columnwidth,keepaspectratio]{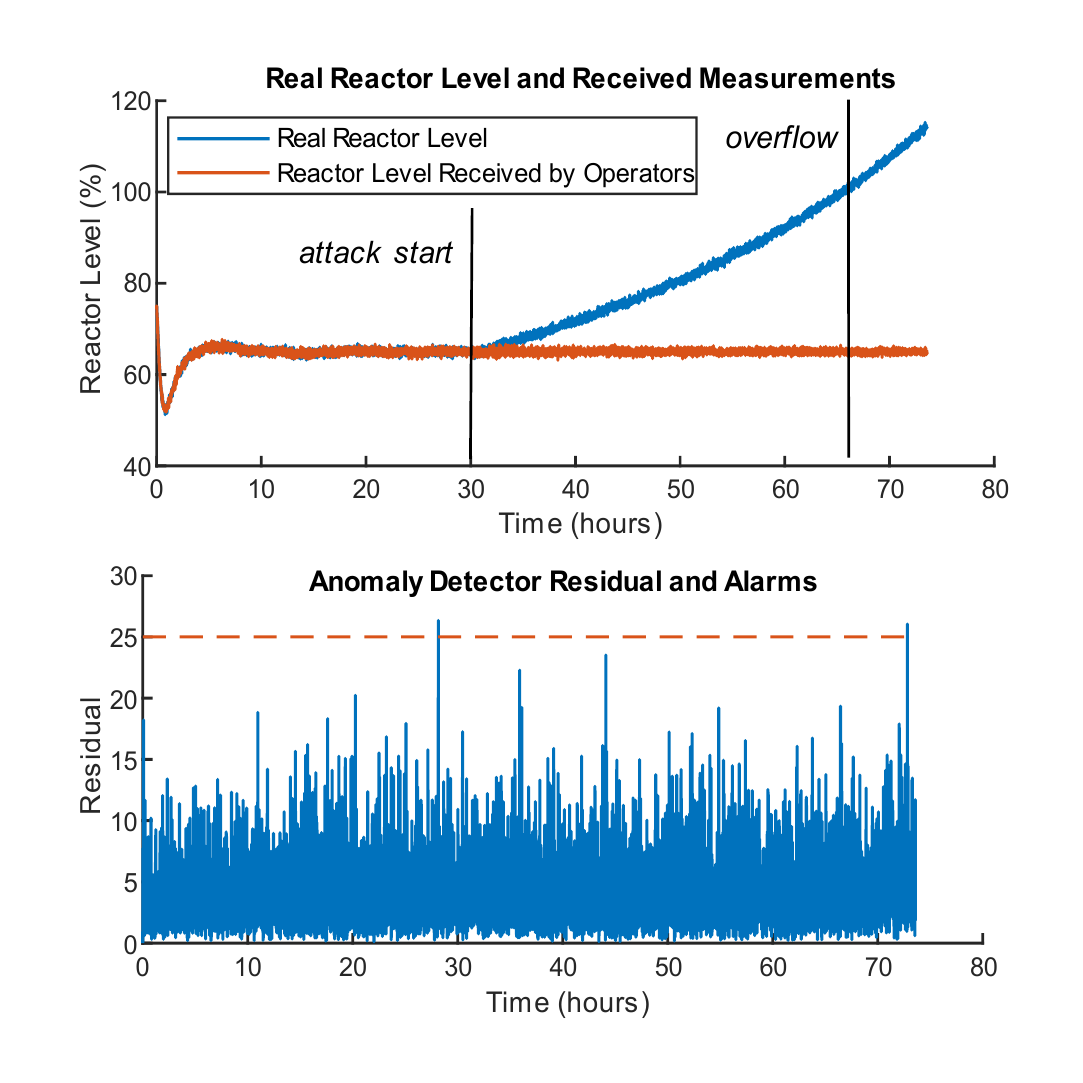}
    \caption{Real and received reactor level values (top), as well as anomaly detector residual metrics (bottom). The reactor overflows at \(t\approx 67h\), \(34h\) after the start of the attack.}~\label{fig:example.plots}
\end{figure}

The attacker takes advantage of the safety-critical operating mode of the reactor, and
modifies level sensor values such that the bias between real and
received values grows slowly over time. This in turn tricks the
controller into slowly increasing the level in the reactor,
which is driven to the point of overflow (top part of Figure~\ref{fig:example.plots}). This may have significant consequences, such as a
fire, especially if the reactor is operating at high temperature or pressure. In addition, data stored on low-level control devices
(e.g. Programmable Logic Controllers' (PLC) configurations,
sensor/actuator states) that could be useful during incident
response, would be lost.\@
\par
\par
In this example, the system is equipped with a chi-squared anomaly detector commonly used to detect deviations from normality which can be caused by system faults or attacks. At each time step, the anomaly detector uses control inputs, historical sensor measurements, and a model of the system to predict sensor measurements. These predictions are then compared with the received measurements (red line in the top of Figure~\ref{fig:example.plots}) using a metric, called residual. The residual computes the difference between expected and received measurements and uses a statistical change detection technique (chi-squares) to detect anomalies once a threshold is crossed. This metric fails to raise any alarm before the reactor level crosses the safety limits. It is worth noting that some data-driven anomaly detectors such as the one proposed by Aoudi et al.~\cite{Aoudi_Truth_2018} may be able to detect the attack illustrated in this example. However, these ``model-agnostic'' detectors have been shown to still miss certain attacks as they are not able to properly capture the physical properties of the system~\cite{Erba_No_2020}. In the present work, we consider model-based anomaly detectors, and particularly the chi-squared anomaly detector as it is widely studied in the literature.
\par
Existing online monitoring techniques which rely on a notion of proximity to unsafe states (e.g.~\cite{Carcano_multidimensional_2011,Castellanos_modular_2019}) may not be able to detect the illustrated attack since they rely on raw sensor values to measure a distance metric to unsafe states. In the case of the present example, the attacker has forced the received sensor values to appear lower than their real counterparts. Therefore, the evolution of the system towards an unsafe state will not be obvious if the proximity measure relies on raw sensor values. 
\par
Differently from previous work, we monitor whether a potential
stealthy attack tampering with control devices can take the system to an unsafe
operating region. We impose on the attacker constraints brought by the anomaly detector and the physics underlying the system, to check whether an attack can damage the system before being detected. Our framework triggers an early warning when a measure of the likelihood of success of a
stealthy attack in real-time exceeds a given threshold.  Thus, an EWS configured using our framework  would have raised a warning well before the stealthy attack exemplified in this section could cause harm.
%Instead of relying on raw sensor values to monitor the proximity of the system to unsafe states,  Even if raw sensor values do not reflect the real state of the system, the ability of an attacker to actually cause damage without getting detected is constrained 
\par
Early warnings can trigger data collection activities,
which can help profile a potential intrusion and prevent the loss of
potential evidence. Operators can also engage safety measures to prevent harm. However, in this paper we focus on providing ``physics-based'' early warnings and defer a detailed treatment of post-warning measures to future work.
%%% Local Variables:
%%% mode: latex
%%% TeX-master: "main"
%%% End:

%% file: section4.Contribution.tex
\section{Proposed EWS Framework}\label{section:contribution}
%\begingroup
In this section, we present the main contribution of this paper, which is a framework for physics-based EWS in ICS. Our framework builds on the tiered approach for EWS used by Bradford et al.~\cite{Bradford_Towards_2004} and Chivers et al.~\cite{Chivers_Accumulating_2009}. These approaches monitor preliminary indicators, often called \textit{weak evidence} until their evidentiary weight crosses a certain threshold and triggers a warning. The main novelty of the present work lies in its consideration of stealthy attacks on ICS that affect the physical process. To this end, our framework accumulates weak evidence collected by monitoring the physical processes in ICS. In the following, we detail the nature of this evidence and the structure of our framework.
%We consider in this work weak evidence collected by monitoring the physical processes in ICS.\@ Our framework accumulates such evidence as part of a more comprehensive EWS.\@ In the following, we detail the nature of this evidence and the structure of our framework.

\subsection{Framework Structure}
%\begingroup
\begin{figure*}[!t]
    \centering
    \includegraphics[width=\textwidth{},keepaspectratio]{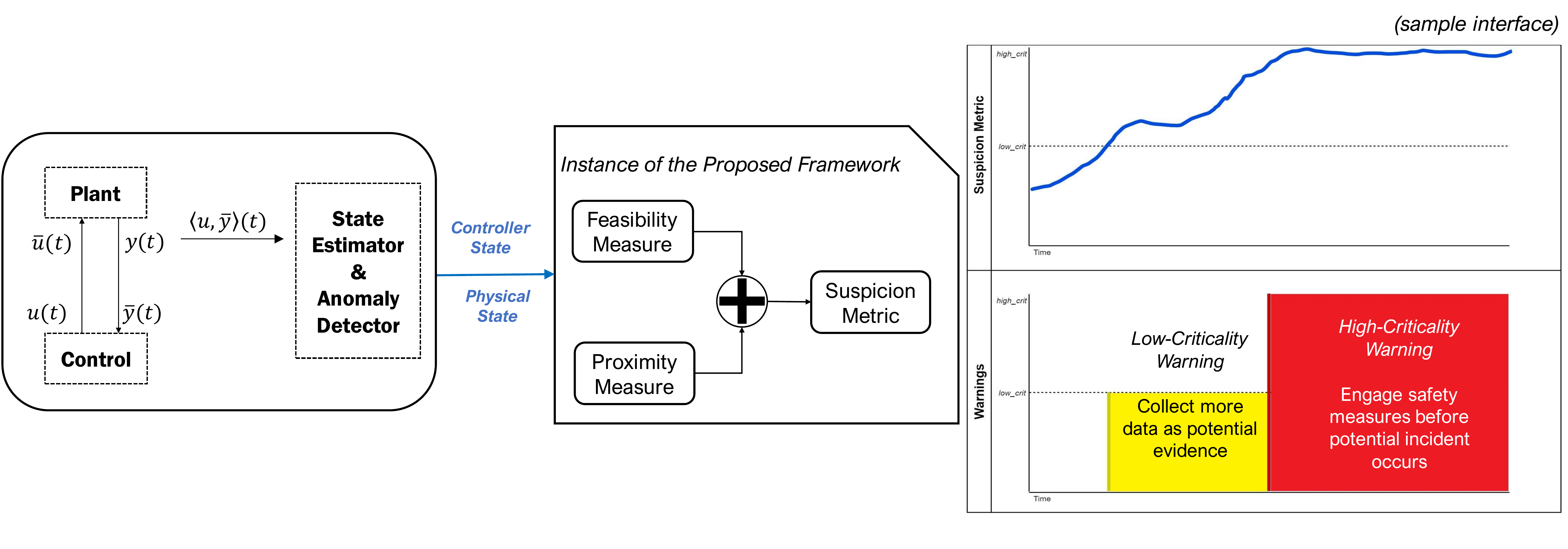}
    \caption{An instance of the proposed framework within a typical control system and a sample interface. \textit{(\(y(t)=\) sensor measurements, \(u(t)=\) control signal, \(\Bar{y}(t) = \) sensor measurements received by the controller, \(\Bar{u}(t)=\) control signal received by system --- respectively at time \(t\).)}}~\label{fig:contribution.fw_positioning}
\end{figure*}
Figure~\ref{fig:contribution.fw_positioning} shows an instantiation  of our framework to a control system. The latter mainly takes as input estimates of physical state variables provided by some state estimator as well as the current state of the controller(s). These estimates are then used to measure the feasibility and proximity grounds for suspicion, which act as weak evidence of a stealthy attack by reflecting its likelihood of success. These two grounds are then combined and measured via a suspicion metric, which in turn reflects their evidentiary weight, and triggers a warning when crossing a certain threshold. Depending on the criticality of the crossed threshold, different actions may follow, such as further evidence collection or safety measures initiation. Identification of these actions is outside the scope of this work. 
%While techniques for evidence collection and safety measures are outside the scope of this work, we provide details on the structure of the framework next. 
\par
\subsubsection{Grounds for Suspicion.}
An attacker wishing to avoid detection will manipulate the system in a way that keeps the difference between estimated predictions and actual sensor readings sufficiently small. Our framework does not attempt to distinguish an anomalous behaviour by comparing the two. Instead, it measures the following \emph{grounds for suspicion}: 
\begin{itemize}
    \item \textit{Feasibility of a Stealthy Attack}: given the dynamics of the system and the constraints imposed by the anomaly detector, we check whether the current state of the system can be taken to an unsafe state (hence causing damage) while avoiding any alarm in the process.
    \item \textit{Proximity to Unsafe States}: given the current state of the system, if an attack is actually taking place without having been detected, we monitor how far would the system be from reaching an unsafe operating region. The closer the system is to this region, the more likely a stealthy attack is successful, as the attack may take less time to achieve its goal.
\end{itemize}
The feasibility of a stealthy attack and the proximity of the system to an unsafe state do not necessarily imply that a malicious activity is taking place. However, they can indicate that a stealthy attack may successfully damage the system. Hence, we consider them to be weak evidence of a potential stealthy attack, in the same manner a failed login attempt may be suspicious but cannot be used as evidence of an intrusion. %As such cues raise suspicions, we refer to them as \textit{grounds for suspicion}.
\par
\subsubsection{Suspicion Metric}
%As it may be infeasible to store raw events that can be used for early warnings, their evidentiary weight is typically reflected in a certain metric. 
The evidentiary weight of events that can trigger early warnings is typically measured using a certain metric.
For example, Chivers et al.~\cite{Chivers_Accumulating_2009} assign a Bayesian score to network nodes that generate events considered as weak evidence. We propose an analogous score, called \textit{suspicion metric}. This metric essentially measures in real-time the likelihood that a potential stealthy attack will cause damage to the system before being detected by combining these two grounds for suspicion. At runtime, a human operator would be provided with the evolution of the suspicion metric over time (Figure~\ref{fig:contribution.fw_positioning}). If the metric crosses a certain threshold, a warning is raised. %This probability can be used to compute the \textit{expected loss} due to a successful stealthy attack, where the expected loss is the sum of the value of all possible losses each multiplied by the probability of that loss occurring. In this work, we assume that all possible losses (i.e.\ damaged assets) have the same value and we focus on the probability of damage caused by a stealthy attack. The case where different assets have different priorities or values is beyond the scope of the current work.
\par
%To design a suspicion metric, we use the system's mathematical model along with analysis tools from the growing literature on CPS security to provide a measure combining feasibility of stealthy attacks and proximity to unsafe operating regions. %This measure then provides a likelihood of damage by a stealthy attack b.
\par
 %followed by certain actions depending on the criticality of the warning --- such as further evidence collection or engagement of safety measures.
%\endgroup

\subsection{Suspicion Metric Soundness Principles}
%\begingroup
As several types of physical systems can be modelled with different  formalisms (e.g.\ continuous-state vs.\ discrete-event/hybrid), we do not attempt to propose a formula to compute a suspicion metric. Instead, we provide \textit{soundness principles}, so that irrespective of the system where the framework is  instantiated, the metric can reflect the grounds for suspicion in a sound manner:
\begin{enumerate}
    \item The metric must include at least one clear threshold which if crossed, a warning is issued. We propose two thresholds (Figure~\ref{fig:contribution.fw_positioning}): (i) one of \textit{low-criticality}, which may trigger intensive data collection to proactively check for intrusions; and another (ii) of \textit{high-criticality} typically triggering measures preventing a safety incident.
    \item The metric should increase over a certain time interval if \begin{enumerate*} \item the likelihood of the real physical state of the system diverging from the provided estimate and evolving into an unsafe operating region is increasing (feasibility);  or \item if the system is evolving closer to unsafe states meaning that a potential attack is less and less time consuming for the attacker (proximity).\end{enumerate*}
\end{enumerate}
The first principle ensures that the EWS can advise an operator about the current security situation. The second ensures that the metric provides a measures the evidentiary weight of the grounds for suspicion and can inform operators about the likelihood of success of a potential stealthy attack.
%\endgroup

\subsection{Framework Configuration Requirements}
%\begingroup
The configuration of the framework involves mainly providing measures of feasibility and proximity to construct a suspicion metric according to the soundness principles provided earlier. This can be performed by reusing existing techniques proposed in the domain CPS/ICS security. For example, techniques to compute reachable sets under a given stealthy attack~\cite{Murguia_reachable_2018,Kwon_Reachability_2018} can be used to measure  feasibility, while distance metrics such as the Euclidean or Hamming distance, can be used to measure proximity given the real-valued nature of most physical state variables.
\par
Therefore, to configure the framework, we require mainly (i) a mathematical model of the system, including its controller and the unsafe operating region; (ii) information on the used anomaly detection method; and (iii) a model of an attack at the level of physical process. Note that (i) is standard in control engineering and safety analysis, while the threat model (iii) is only required to show the effect of a potential attack on the control loop. There exist several works~\cite{Huang_Understanding_2009,Teixeira_Secure_2015a} that provide effective ways of modelling such effects.
\par
In this paper we focus on applying the framework to systems that can be modelled using the Linear-Time Invariant (LTI) modelling framework --- we defer the study of other systems to future work. Given a certain operating range, several control systems can be approximated with high accuracy by an LTI model using well-established techniques. This modelling framework is especially applicable to several problems in the process control industry~\cite{Polisetty_Error_2019}.
\par
\begin{figure*}[!t]
    \centering
    \includegraphics[width=\textwidth{},keepaspectratio]{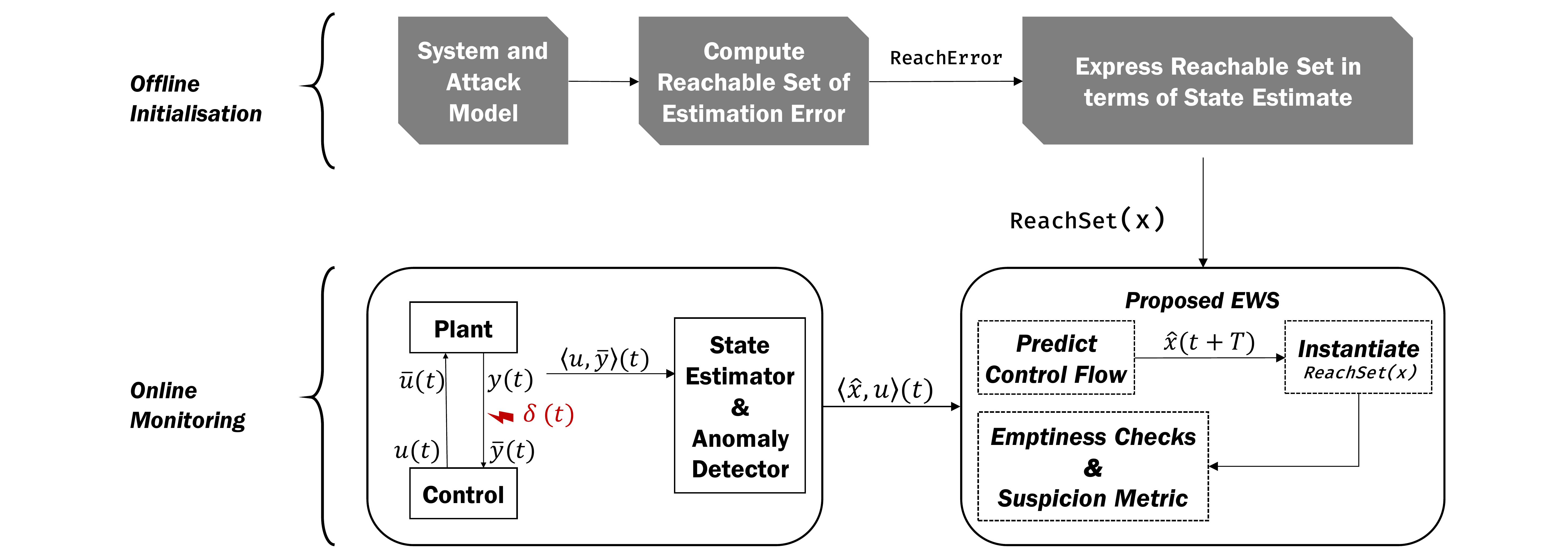}
    \caption{Proposed instantiation of the framework to LTI systems. (\textit{\(\delta(t)\) denotes potential attack signal.})}~\label{fig:framework_instantiation}
\end{figure*}
The proposed instantiation of the framework to LTI systems is outlined in Figure~\ref{fig:framework_instantiation}. We formulate the suspicion metric computation as a reachability problem. Namely, given the real-time estimated physical state of the system, the reachability problem asks whether a stealthy attack can cause damage to the system without being detected. To enable efficient reachability analysis in real-time, our approach computes lightweight symbolic ellipsoidal approximation of the reachable set under attack offline, thus restricting the bulk of the computation to a design-time activity. This is possible by considering the evolution of the state estimation error under stealthy attacks rather than the physical state itself. By using analysis tools from the literature, namely the method developed by Murguia et al.~\cite{Murguia_reachable_2018}, we obtain an approximation of the reachable set of the error in the form of an ellipsoid centred at a given state estimate. The real state of the system, if it is under a stealthy attack, lies in this ellipsoid. The measures of feasibility and proximity subsequently rely on checking whether this set intersects a predefined set of unsafe states. To make these emptiness checks possible in real-time, we take advantage of the ellipsoid nature of the reachable set approximation and the fact that in most scenarios, safety constraints can be interpreted geometrically as a union of half-spaces. In this case, emptiness checks reduce to checking the sign of the distance between the ellipsoid and the half-spaces composing the unsafe set.
\par
%\endgroup

%\endgroup
%%% Local Variables:
%%% mode: latex
%%% TeX-master: "main"
%%% End:

%% file: section5.Case_Study.tex
\section{System Description}\label{section:system}
This section describes the modelling formalism used to represent the system and the controller, the anomaly detector and the threat model.

\subsection{Physical System and Controller}
%\begingroup
The evolution of the physical state of a standard, frequently used Linear-Time Invariant (LTI) model is given as follows:
\begin{small}
\begin{equation}\label{eq:sys.ltimodel}
  \left \{
    \begin{aligned}
      x(k+1) &= Ax(k) + Bu(k) + w(k)\\
      y(k) &= Cx(k) + v(k)
    \end{aligned}
  \right.
\end{equation}
\end{small}
The matrices \(A,\;B,\;\text{and} \;C\) are real, time-invariant, and of appropriate dimensions. The state of the system is given by the vector \(x(k) \in \R^n\), sensor measurements by \(y(k) \in \R^m\)  and control input by \(u(k) \in \R^p\) where \(k = t/\Delta_t\in \mathbb{N}\) denotes discrete time instants with \(\Delta_t\) being the sampling period. Process disturbances \(w(k)\) and sensor noise \(v(k)\) are assumed to follow a zero-mean Gaussian distribution with covariance matrices \(\Sigma_1\) and \(\Sigma_2\), respectively. We assume that the system is observable and controllable in a control-theoretical sense. Furthermore, the system is equipped with an output feedback control loop, such that given received sensor measurements \(\bar{y}(k)\) and a set-point reference \(y_r(k)\), a control signal \(u(k) = \mathcal{K}[\Bar{y}(k)-y_r(k)]\) based on the control law \(\mathcal{K}[.]\) is sent to the process at each time step. We assume that the system is stabilised by this controller.
\par
A subset of state variables, denoted as ``critical'', and grouped in a vector \(x_c = C_c x\), \(x_c\in\R^n_c\), \(C_c\in\R^{n_c\times n}\), are required to remain within certain bounds to ensure safe operation. Unsafe conditions can be written in the form of a linear combination of the state variables. Thus each linear combination of state variables denoting safety constraints can be geometrically interpreted as a half-space in \(\R^n\). Let \(\mathcal{S}_u\) denote the unsafe operating region, this can in turn be interpreted as a union of half-spaces as follows:

\begin{small}
\begin{equation}\label{eq:sys.unsafeset}
  \mathcal{S}_u = \left \{x(k)\in\R^n\;|\; \bigcup_{i=0}^{n_c} C_{c,i} x(k) \geq b_i \right \}
\end{equation}
\end{small}
Where \(b_i\) denotes the safety bound on the \(i\)\textsuperscript{th} critical state variable (or the \(i\)\textsuperscript{th} half-space scalar from a geometric point of view), and \(C_{c,i}\) denotes the \(i\)\textsuperscript{th} row of the matrix \(C_c\).
\par
Although our main concern in this paper is with stealthy attacks that seek to cause physical damage to the system, our modelling framework can accommodate other objectives for stealthy attacks. For example, if we are worried about attackers causing economic loss by driving the system to an ``expensive'' operating state, then the relevant state variables and constraints can be added to \(\mathcal{S_u}\) to express such an expensive operating region.
%\endgroup
\subsection{Anomaly Detector}
%\begingroup
At a time \(k\), given previous estimates and control actions, the state estimate \(\Hat{x}(k)\) and expected sensor measurements \(\Hat{y}(k)\) are provided by a Kalman filter:

\begin{small}
\begin{equation}\label{eq:sys.kalman}
  \left \{
  \begin{aligned}
      \Hat{x}(k) =& \; A\Hat{x}(k-1) + Bu(k-1) 
               \\ &+ L(\Bar{y}(k-1) - C\Hat{x}(k-1))\\
      \Hat{y}(k) =& \; C\Hat{x}(k)
  \end{aligned}
  \right.
\end{equation}
\end{small}
Where the design parameter \(L\) is the observer gain matrix, the existence of which is guaranteed by the observability of the system. The estimated sensor measurement \(\Hat{y}(k)\) is compared with the received value \(y(k)\) using a residual metric; \(r(k) := y(k) - \Hat{y}(k)\). Under nominal conditions, the residual metric has a zero-mean and a covariance matrix \(\Sigma \). To check for this hypothesis, a chi-squared metric, \(z(k) = r^T(k)\Sigma^{-1}r(k)\) is computed and compared with a threshold \(\tau \), such that exceeding this threshold implies a possible anomaly and raises an alarm. \(\tau \) can be set according to a desired false alarm rate \(\beta \). The reader is referred to~\cite{Murguia_reachable_2018} for more detail on the derivation of the observer gain matrix \(L\), the anomaly detection threshold \(\tau \), and the residual's covariance matrix \(\Sigma \) under nominal conditions.
\par
%\endgroup
\subsection{Threat Model}
%\begingroup
In this paper we consider \textit{false data injection} attacks on sensors, which consist of falsifying sensor values such that the controller drives the system into unsafe operating levels. Such attacks are typically modelled as a bias imposed on \(y(k)\)~\cite{Teixeira_Secure_2015a}. Let \( \{k_s,\dots, k_f\} \) be the time period of the attack, the actual sensor readings \(\Bar{y}(k)\) received by the controller are then given as:

\begin{small}
\begin{equation}\label{eq:sys.threatModel}
    \Bar{y}(k) =
    \left \{
      \begin{aligned}
        & y(k) + \delta(k)\; \forall k\in \{k_s,\dots,k_f\};\\
        & y(k)\; \text{otherwise}
      \end{aligned}
    \right.
\end{equation}
\end{small}
\par
Under such attack, the anomaly detector's chi-squared metric is given by:

\begin{small}
\begin{equation}\label{eq:threatModel_detectmetric}
  z(k) = (y(k) - \Hat{y}(k) + \delta(k))\Sigma^{-1}(y(k) - \Hat{y}(k) + \delta(k))
\end{equation}
\end{small}
The attacker, having knowledge of the anomaly detector parameters (i.e. \(\Sigma \), \(\beta \) and \(\tau \)) can maintain the stealthiness of the attack by ensuring that \(\delta(k)\) maintains a nominal false alarm rate; i.e. \(\Pr[z(k) \leq \tau] = 1-\beta \). We use this characterisation of a stealthy attack because it represents more realistically an advanced attacker wishing to remain stealthy until at least achieving the objective of damaging the system. %Under nominal operation, the anomaly detector will raise alarms due to its false alarm rate. A sudden disappearance of these alarms in practice may raise suspicions in operators about a potential malfunction or intrusion. This may lead eventually to the detection of the attack.
%\par
In practice, given \(K\) time steps, the attacker may choose to raise alarms for \(\beta K\) steps; thus mimicking the false alarm rate as closely as possible. Due to space limitations, the reader is referred to~\cite{Hashemi_comparison_2018} for a more detailed description and analysis of the distribution of the detector metric under such attack.
%\endgroup

%%% Local Variables:
%%% mode: latex
%%% TeX-master: "main"
%%% End:

%% file: section6.Instantiating.tex
\section{Instantiating the Proposed Framework for LTI Systems}\label{section:instance}

The proposed physics-based EWS component takes as input the vector of estimated state variables \(\Hat{x} \) in addition to the state of the controller. We initialise the suspicion metric \(\mathbf{SUSP}\) as a function of two terms: feasibility \(\mathbf{FEAS}\) and proximity \(\mathbf{PROX}\). In this section, we instantiate the proposed framework for the system described in Section~\ref{section:contribution}. We use analysis tools from the literature to construct a formula for \(\mathbf{FEAS}\) and \(\mathbf{PROX}\). In a previous work~\cite{Azzam_Efficient_2021}, we showed how to perform efficient online safety checking for LTI systems under stealthy attacks. In this section, we describe the safety checking algorithm and we extend it by including the suspicion metric and designing thresholds for warnings of different criticality in order to fit the algorithm into the proposed framework. We also show how the proposed algorithm satisfies the soundness principles proposed in Section~\ref{section:contribution}.

\subsection{Feasibility Measure}
%\begingroup
The attack in~(\ref{eq:sys.threatModel}) is defined to be feasible if (i) the corrupting signal can maintain the nominal false alarm rate throughout the attack and (ii) at the end of the attack the system can be driven to unsafe operation. This can be stated as a reachability problem. Namely, let \(\mathcal{R}_x(k)\) be the set of reachable states at time \(k\) due to the attack (\ref{eq:sys.threatModel}):
\begin{small}
\begin{multline}\label{eq:instance.reachSet}
  \mathcal{R}_x(k) = \{ x(k)\in \R^n \;|\; x(k)\; \text{is s.t.}
  \; (\ref{eq:sys.ltimodel}) \\ \land \delta(k)\;\text{is s.t.}\;
  \Pr[z(k) \leq \tau] = 1-\beta \}
\end{multline}
\end{small}
The attack (\ref{eq:sys.threatModel}) is then feasible at time \(k\) if, for the next \(K\) time instants, there exists an instant \(k_f\in \{k,\dots,k+K\} \) such that \(\mathcal{R}_x(k_f)\cap \mathcal{S}_u \neq \emptyset \). As a measure of feasibility, we propose to use a function of the size of this intersection. However, computing \(\mathcal{R}_x(k)\) and the size of \(\mathcal{R}_x(k_f) \cap \mathcal{S}_u\) exactly in real-time is intractable. Furthermore, since \(x(k)\) is partially driven by the Gaussian noise \(w(k)\) which has an infinite support, computing \(\mathcal{R}_x\) using deterministic methods will yield an unbounded set.
\par
We address intractability by approximating a symbolic reachable set offline parametrised by the state estimate. We compute the resulting set under the assumption of a bound on the energy of the noise with a certain confidence level, thus preventing unbounded reachable sets. As such, computing the a feasibility measure (and subsequently the suspicion metric) involves an offline initialisation step as well as online emptiness checks of \(\mathcal{R}_x(k_f) \cap \mathcal{S}_u\). %We detail these procedures in the following.
\par
\subsubsection{Offline Initialisation}
To approximate a symbolic reachable set offline parametrised by the current state estimate, we consider the reachable set \(\mathcal{R}_e\) of the estimation error \(e(k):= x(k) - \Hat{x}(k)\) under an attack. Assuming the initial error at the beginning of an attack is always almost zero, this set is independent of the physical state at the start of the attack. Hence, computing it offline would provide a symbolic reachable set as a function of the provided state estimate, which can then be instantiated online. We use the method proposed by Murguia et al.~\cite{Murguia_reachable_2018} to compute an ellipsoidal approximation of the reachable set of estimation error under a stealthy attack.
\par
Based on Equation (\ref{eq:sys.kalman}), the estimation error under an attack evolves according to the following:

\begin{small}
\begin{equation}\label{eq:estimationerror}
  e(k+1) = Ae(k) - L(y(k) - \Bar{y}(k) + \delta(k)) + w(k)
\end{equation}
\end{small}
To address the problem of computing \(\mathcal{R}_e\) when the error is partially driven by the Gaussian noise \(w(k)\) and the attack-dependent sequence \(\Bar{\delta}(k) = y(k) - \Bar{y}(k) + \delta(k) \), we set a confidence level on the energy of both of these vectors. Given the threat model described in (\ref{eq:sys.threatModel}), we have for the sequence \(\bar{\delta}(k)\) that  \(\Pr[z(k)\leq \tau] = \Pr[\lVert \Sigma^{-1/2}\Bar{\delta}(k) \rVert^2\leq \tau]= 1-\beta \) where \(\lVert . \rVert \) denotes the \(L_2\)-norm. As for the noise, since it follows a Gaussian distribution, a bound \(\Bar{w}\) on its energy \(\lVert w(k)\rVert^2\) can be set for a desired confidence level \(p = \Pr[\lVert w(k)\rVert^2 \leq \Bar{w}]\) using the gamma or the chi-squared distribution~\cite{Murguia_reachable_2018,Hashemi_comparison_2018}.
\par
Using this truncation of the distribution of \(\Bar{\delta}(k)\) and \(w(k)\), we compute an ellipsoidal approximation \(\mathcal{E}^p_e\) of \(\mathcal{R}^p_e\) offline for a desired confidence level \(p\). A larger confidence level would lead to a larger set, at the cost of being overly conservative with the emptiness checks. A reasonable choice for \(p\) would be \(1-\beta \), as the false alarm \(\beta \) is designed to be small. This also simplifies the computation of the reachable set, since for \(p=1-\beta \), we readily have \(\Pr[\lVert w(k)\rVert^2 \leq \Bar{w}] = \Pr[z(k)\leq\tau]\) under the attack in (\ref{eq:sys.threatModel}).
\par
Given the system model (\ref{eq:sys.ltimodel}), the Kalman gain \(L\), the anomaly detector threshold \(\tau \), and the \(p\)-probable bound \(\Bar{w}\) on the process noise energy, it is possible to compute an ellipsoidal approximation \(\mathcal{E}_e^p \supseteq \mathcal{R}_e^p\) of the following form:

\begin{small}
\begin{equation}
  \mathcal{R}_e^p \subseteq \mathcal{E}_e^p = \{e(k)\;|\; e^T(k)\mathbf{\Pi}^{-1}e(k) \leq 1\}
\end{equation}
\end{small}
Where \(\mathbf{\Pi} \) is called the ellipsoid's shape matrix. The computation of \(\mathbf{\Pi} \) involves solving a Linear Matrix Inequality (LMI) problem given the aforementioned parameters. Note that since we assume the system to be stable, the matrix \(\mathbf{\Pi}\) exists~\cite{Murguia_reachable_2018}. Due to space limitations, the reader is referred to~\cite{Murguia_reachable_2018} and~\cite{Hashemi_comparison_2018} for more details on this procedure and the effect of the choice of \(p\) on the tightness of the ellipsoidal approximation.
\par
Note that this step is performed only once offline, and only the matrix \(\mathbf{\Pi} \) needs to be stored for online emptiness checks. Therefore, the computation of this ellipsoidal approximation does not affect real-time performance. Given the matrix \(\mathbf{\Pi} \), and replacing \(e(k)\) by its definition, we obtain a symbolic ellipsoidal approximation \(\mathcal{E}_x^p(\Hat{x}(k))\) of the reachable set \(\mathcal{R}_x^p(x(k))\) of the actual system state \(x(k)\), parametrised by the current state estimate \(\Hat{x}(k)\):

\begin{small}
\begin{equation}\label{eq:reachsetx}
    \begin{aligned}
    &\mathcal{R}_x^p(x(k)) \subseteq \mathcal{E}_x^p(\Hat{x}(k)) =\\ 
    &\{x(k)\in\R^n\;|\;{(x(k)-\Hat{x}(k))}^T\mathbf{\Pi}^{-1}(x(k)-\Hat{x}(k)) \leq 1\}
    \end{aligned}
\end{equation}
\end{small}
This ellipsoidal approximation can then be instantiated online at a time \(k\) given the current state estimate \(\Hat{x}(k)\).

\subsubsection{Online Emptiness Checks}
\label{sec:EmptinessChecks}
At runtime, given the current physical state estimate and the state of the controller, we predict the state of the system for \(K\) steps into the future using the identified model of the system. For each predicted state \(\Hat{x}(k + l)\), \(l\in \{0,\dots,K\} \), we instantiate the ellipsoidal approximation \(\mathcal{E}_x^p(\Hat{x}(k+l))\) of the reachable set under a potential stealthy attack. Upon encountering a state where \(\mathcal{E}_x^p(\Hat{x}(k+l)) \cap \mathcal{S}_u \neq \emptyset \), the prediction stops, and we compute the size of this intersection as a feasibility measure.
\par
At each predicted state \(\Hat{x}(k + l)\), we take advantage of the ellipsoidal nature of \(\mathcal{E}_x^p(\Hat{x}(k+l))\) and the fact that \(\mathcal{S}_u\) can be interpreted geometrically as a union of half-spaces to perform efficient emptiness checks of their intersection. For each hyperplane delimiting a half-space in \(\mathcal{S}_u\), we compute the distance from the ellipsoid \(\mathcal{E}_x^p(\Hat{x}(k+l))\). The intersection is then non-empty if the distance value is negative~\cite{kurzhanski_ellipsoidal_2000}.
\par
When the intersection is non-empty, it is possible to approximate the intersection of \(\mathcal{E}_x^p(\Hat{x}(k+l))\) with each of the half-spaces \(\mathcal{H}_i \subseteq \mathcal{S}_u\) using an ellipsoid \(\mathcal{E}^p_{x,i}\) of shape matrix \(\mathbf{\Pi}_i \). Boyd and Vandenbergh~\cite{boyd_convex_2004} provide an equation to efficiently compute this shape matrix, which we omit here for brevity. We use the ratio of the volume of this approximate ellipsoid to the reachable ellipsoid to measure feasibility:
\begin{equation}\label{eq:inst.feas}
\begin{small}
  \mathbf{FEAS}(k) = V_{\mathcal{E},\Hat{l}} / V_{\mathcal{E}}
  \end{small}
\end{equation}
Where \(V_{\mathcal{E},\Hat{l}} =  \max_{i=1,\dots,n_c}[\mathbf{vol}(\mathcal{E}^p_{x,i})]\) is the maximum intersection volume obtained at time \(k + \Hat{l}\) among the intersections of \(\mathcal{E}_x^p(\Hat{x}(k+l))\) with each of the half-spaces \(\mathcal{H}_i \subseteq \mathcal{S}_u\).  \(V_{\mathcal{E}} = \mathbf{vol}(\mathcal{E}^p_x) \) is the volume of the reachable ellipsoid.

%\endgroup

\subsection{Proximity to Unsafe States}
%\begingroup
Given the real-valued nature of the physical state variables, one can employ a simple measure based on Euclidean distance to compute the proximity of the system to the set of unsafe states. 
This approach was employed by Coletta et al.~\cite{Coletta_Predictive_2018}. However, as the actual state of the system may be different from the given estimate (due to a potential stealthy attack), this simple distance measure may not reflect the actual proximity of the system to unsafe operating region. Such measure also does not reflect how fast the system may evolve to an unsafe state under a potential stealthy attack.
\par
We make use of the procedure used to compute the symbolic reachable set explained in Section~\ref{sec:EmptinessChecks}. If we find that \(\mathcal{E}_x^p(\Hat{x}(k+l)) \cap \mathcal{S}_u \neq \emptyset \) for an \(l = \Hat{l}\), then we can conclude that under a potential stealthy attack, the system may be damaged after at least \(\Hat{l}\) time instants. Hence, we use the following as a measure of proximity:
\begin{equation}\label{eq:inst.prox}
\begin{small}
  \mathbf{PROX}(k) = 1/(1+\Hat{l})
  \end{small}
\end{equation}

\subsection{Algorithm and Metric Soundness}
%\begingroup

\begin{algorithm}[!t]
\begin{small}
  \caption{Computing the Suspicion Metric Online}\label{alg:online}
  \begin{algorithmic}[1]
  \Statex{\textsc{Inputs:} (\(K, \mathbf{\Pi} ,\Hat{x}(k)\), ControlState, \(\mathcal{S}_u\))}
  \Statep{\(\Hat{x}_p \gets \Hat{x}(k)\)}
  \ForAll{l \(\in \{0,\dots,K\} \)}
  \Commentl{Instantiate the reachable ellipsoid at current predicted state}
  \Statep{ReachEll \(\gets \)\Call{Ellipsoid}{\(\Hat{x}_p\),\(\mathbf{\Pi} \)}}
    \ForAll{\(\mathcal{H}_{p,i} \subset \mathcal{S}_u \)}
      \Statep{DistToUnsafe \(\gets \) \Call{dist}{ReachEll,Hyperplane}}
      \If{DistToUnsafe \(< 0\)}
        \Commentl{Raise non-empty intersection flag and compute the shape matrix of the intersection ellipsoid}
        \Statep{isNonEmpty\(\gets \) \False}
        \Statep{\(\mathbf{\Pi}_i\) \(\gets \) \Call{EllIntersect}{ReachEll,Hyperplane}}
        \Statep{\(V_{\mathcal{E},i}\gets \) \Call{Volume}{\(\mathbf{\Pi}_i \)}}
      \Else{}
        \Statep{\(V_{\mathcal{E},i}\gets \) 0} 
      \EndIf{}
    \EndFor{}
    \If{!isEmpty}
    \Commentl{Break the loop and return the suspicion metric}
      \Statep{\textbf{FEAS} \(\gets \)\Call{Max}{\(V_{\mathcal{E},i}\)}/\Call{Volume}{ReachEll}}
      \Statep{\textbf{PROX} \(\gets 1/(1+l)\)}
      \Statep{\textbf{SUSP} \(\gets \) \textbf{FEAS \(\times \) PROX}} 
      \Statep{\textbf{return} \textbf{SUSP}}
    \Else{}
      \Commentl{Predict next state}
      \Statep{\(\Hat{x}_p\) \( \gets \) \Call{PredictControlFlow}{\(\Hat{x}_p\), ControlState}}
    \EndIf{}
  \EndFor{}
  \Commentl{If no intersection with the unsafe set is found to be non-empty, then the suspicion metric is 0}
  \Statep{\textbf{return} 0}
  \end{algorithmic}
  \end{small}
\end{algorithm}

Algorithm 1 outlines the steps taken online to perform online safety checks and compute the suspicion metric. The computation of this metric consists of three main steps: (1) Given the current estimated state \(\Hat{x}(k)\) of the system and the state of controllers, we predict the evolution of the state for a specified number of time steps into the future. (2) For each predicted state, the approximate reachable set under stealthy attacks is instantiated and emptiness checks of its intersection with the set of unsafe states are performed. (3) If the intersection is non-empty at a time \(k+\Hat{l}\), the prediction stops, and the suspicion metric is computed as follows:
\begin{equation}\label{eq:inst.susp}
\begin{small}
  \mathbf{SUSP}(k)= \mathbf{FEAS}\times\mathbf{PROX} = \frac{V_{\mathcal{E},\Hat{l}}}{V_{\mathcal{E}}(1+\Hat{l})}
  \end{small}
\end{equation}
If the intersection is empty for all the states predicted within the specified number of steps, then \(\mathbf{SUSP}(k) = 0\).
\par
\subsubsection{Algorithm Complexity}
In Algorithm 1, the prediction of the state of the system relies on an identified LTI model. As evident from Equation (\ref{eq:sys.ltimodel}), the computation of the next state involves mainly vector addition and matrix-vector multiplication --- operations that scale polynomially with the number of physical states. Thus, given a fixed number of physical states, the prediction is expected to scale linearly with the maximum number of time steps for prediction \(K\). Furthermore, checking the emptiness of the intersection of the current reachable ellipsoid with the set of unsafe states relies on computing the distance between the two sets. This distance, whose formula can be found in~\cite{kurzhanski_ellipsoidal_2000}, relies also on performing matrix-vector multiplication and computing vector norms, which scales polynomially with the number of states. For a fixed number of states, and since this distance is computed for each safety condition in the set of unsafe states, the emptiness checks will scale linearly with the number of safety constraints. The same reasoning applies to the matrix intersection procedure~\cite{boyd_convex_2004} and its corresponding volume, which involves matrix addition and determinant operations. Several constant parameters involved in the computation of ellipsoid-to-half-space distances, the intersection between the two sets, and the feasibility metric can be pre-computed offline to improve real-time performance. These parameters include the volume of the pre-computed reachable ellipsoid and the norms of the half-space normal vectors \(C_{c,i}\) (Equation (\ref{eq:sys.unsafeset})) representing safety conditions.
\subsubsection{Metric Soundness}
An increase in the value of the metric can imply one of the following: (i) \(V_{\mathcal{E},\Hat{l}}\) is increasing, indicating that it is becoming increasingly likely for the actual state of the system to diverge from the estimate and enter an unsafe operating region due to a stealthy attack; (ii) \(\Hat{l}\) is decreasing, indicating that the system is in increasing proximity to unsafe operation regions. Thus, the proposed metric serves as a measure of likelihood of the attacker being able to take the system into unsafe states (feasibility) penalised by the number of time steps required to damage the system (proximity).
\par
%The suspicion metric can complement other preliminary indicators (e.g.\ from cyber intelligence, network events etc.) considered in an EWS. 
To raise physics-based early warnings, we propose two thresholds based on the value of \(\Hat{l}\) returned by Algorithm 1. This value indicates the time that would be needed by a potential attack to cause damage before being detected. Let \(l_1\) be the time required by the operators to perform necessary preemptive actions after a low-criticality warning, and let \(l_2\) be the time necessary to perform potentially more drastic actions after a high-criticality warning, with \(l_2 < l_1\). \(l_1\) and \(l_2\) can be set based on expert knowledge of the system in question and the post-warning measures to be taken.
\par
Accordingly, we set two conditions for different criticality thresholds in this case study:
\begin{enumerate}
  \item A ``low-criticality'' warning is raised when \(\Hat{l} \leq l_1\) and \(V_{\mathcal{E},l_1}/V_{\mathcal{E}} \geq 0.5\). These conditions imply that if the system is under a potential stealthy attack, the damage will likely take place after at most \(l_1\) time instants. However, \(l_1\) is sufficient to perform preemptive low-criticality actions, such as collecting potential evidence of an attack.
  \item A ``high-criticality'' warning is raised if the suspicion metric shows that the attacker is likely to cause damage in a smaller time frame. Namely, this type of warning can be raised when \(\Hat{l} \leq l_1\) and \(V_{\mathcal{E},l_1}/V_{\mathcal{E}} \geq 0.5\) with \(l_2 < l_1\). In this case, high-criticality preemptive actions can be taken, such as engaging safety measures.
\end{enumerate}
\par
The above discussion shows that the proposed metric satisfies the soundness principles proposed in Section~\ref{section:contribution}.
%\endgroup

%%% Local Variables:
%%% mode: latex
%%% TeX-master: "main"
%%% End:

%% file: section7.Evaluation.tex
\section{Evaluation}\label{section:usecase}
In this section, we  describe the virtual testbed that we used to evaluate our framework. Using numerical simulations we validate whether our framework can warn well in advance of damage caused by a potential stealthy attack. We also assess scalability and network performance of our framework. Finally, we  discuss the usefulness of our framework in supporting early evidence collection in a use case scenario. All activities conducted to support the evaluation were performed on  an Intel i7--9750H CPU clocked at 2.6 GHz with 16GB of RAM memory.

\subsection{Virtual Testbed}
%\begingroup
%In this subsection we describe the experimental framework used to validate the framework and assess performance and scalabilitythat we used in validation and evaluation, followed by an overview of our methodology.

%\subsubsection{Experimental Framework}
%\begingroup
To evaluate our framework we rely on a modified simulation of the Tennessee-Eastman Process (TEP)~\cite{Downs_plant_1993}. This is a benchmark chemical process which is used extensively to study problems in the process control field~\cite{Ricker_Model_1993}. The process involves  an exothermic reactor and units to separate and purify chemical products. The temperature and pressure inside the reactor are maintained using several control loops~\cite{Ricker_decentralized_1996}. In the context of security, the complexity of this process has allowed simulating the realistic behaviour of physical processes under attack~\cite{Geng_survey_2019,Genge_Physical_2014,Krotofil_Resilience_2013,McEvoy_plant_2011,Cardenas_Attacks_2011}. In addition, the simulation-based TEP allows for a low-cost and safe testing of the effect of attacks on physical operation. 
\par
\begin{figure}[!t]
    \centering
    \includegraphics[width=\columnwidth,keepaspectratio]{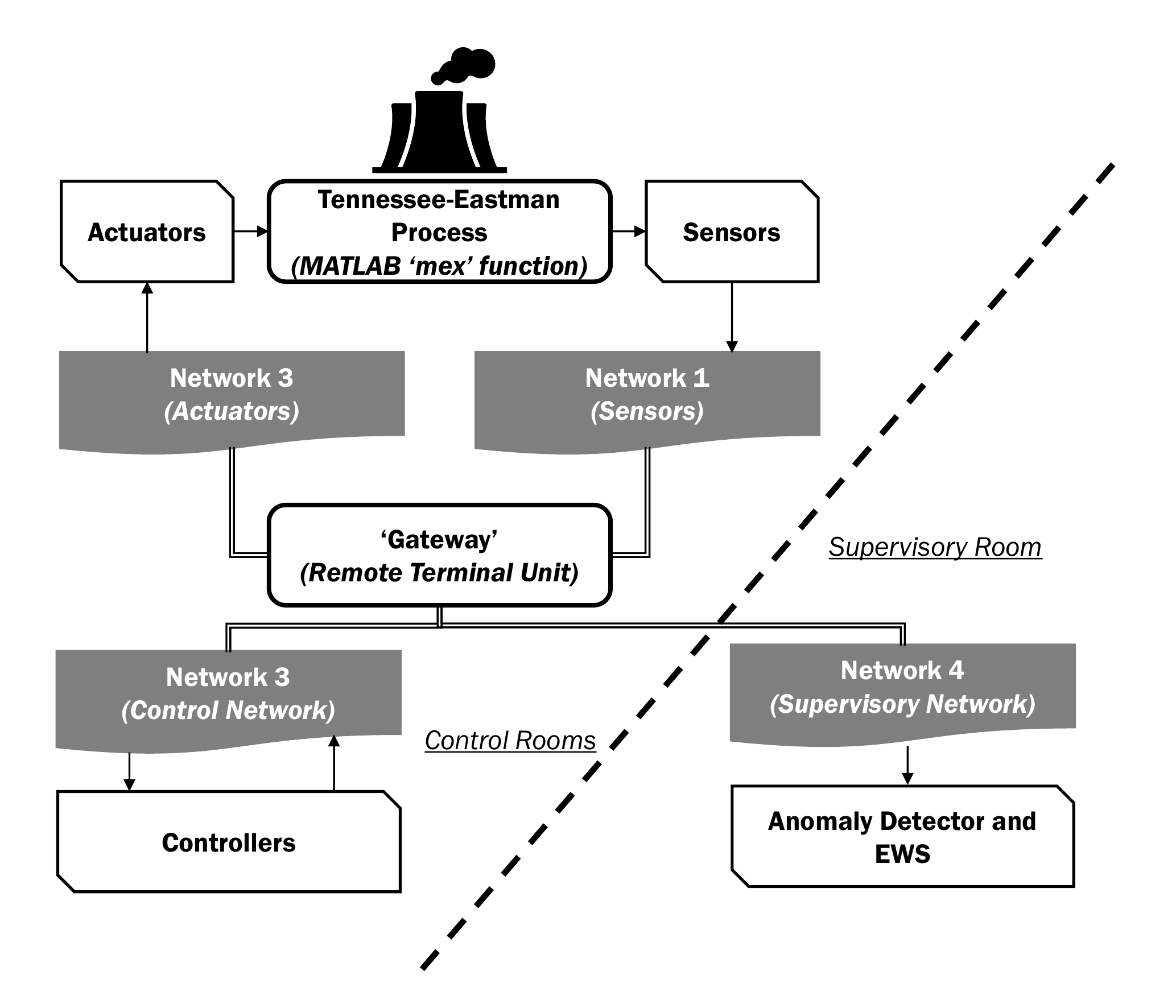}
    \caption{Networked TEP testbed diagram.}\label{fig:testbed}
\end{figure}

We modified the MATLAB/Simulink simulation of the TEP provided by Bathelt et al.~\cite{Bathelt_2015_Revision} by adding blocks to simulate the real-time behaviour of the control network, sensors, actuators, and controllers. We also implemented a Kalman filter-based anomaly detector to estimate process measurements and detect anomalies. A diagram of our testbed is shown in Figure~\ref{fig:testbed}.
\par
The network is divided into four segments connected by a ``gateway'' router to emulate the distributed nature of modern ICS environments. In the control rooms, where the physical process resides, sensors send measurements over the first network segment to the gateway, which forwards them to the appropriate node on the controllers' network. The controllers employ a similar procedure to send control signals to the appropriate actuator nodes. The gateway forwards sensor measurements and controller states to the supervisory control room where the anomaly detector and the proposed instantiated framework (indicated as EWS in Figure~\ref{fig:testbed}) reside. The gateway then emulates Remote Terminal Units (RTU's) which are used to provide an interface between control devices and servers in control rooms.
\par
To simulate the real-time behaviour of sensors, controllers, anomaly detector, and the network we use the MATLAB/Simulink-based TrueTime library~\cite{Cervin_Does_2003}. This library has been adopted  to study the performance of networked control systems~\cite{Kalaivani_Earliest_2019,Brahimi_Comparison_2006}, and also in the context of ICS security~\cite{Farooqui_Cyber_2014}. The TrueTime library provides Simulink blocks to simulate medium access and packet transmission for different industrial network models, such as CAN, Round-Robin, PROFINET, etc. It also provides ``kernel blocks'' for which custom MATLAB or C++ code can be implemented to simulate different nodes (e.g.\ actuators, sensors, controllers etc.) with specified scheduling policies. The TrueTime library simulates medium-access and packet-level network protocols, which are sufficient to  study the overhead incurred by the proposed framework.
%in  .  . We performed these modifications using the TrueTime library~\cite{Cervin_Does_2003}, which is essentially a collection of MATLAB/C++ functions and Simulink blocks designed to simulate real-time control networks and nodes.
\par
%As we are interested only in the effect of attacks on the physical component of an ICS rather than its network, we resort to a simulation of the network rather than a hardware-in-the-loop environment. The objective of simulating the network is to study the overhead incurred by the instantiated framework on the networked performance. In ICS, real-time performance is of utmost importance to ensure the safe and efficient operation of controllers and physical processes. Therefore, any security solution deployed in an ICS environment must not compromise the real-time response of the different control units.

%\endgroup

%\subsubsection{Virtual Testbed Description}
%\begingroup
 We compiled the physical process implemented by Bathelt et al.~\cite{Bathelt_2015_Revision} into a MATLAB ``\texttt{mex} S-function'' to be incorporated in the Simulink-based simulation. We implemented sensor, actuator, and controller codes as TrueTime Kernels, representing the real-time behaviour of control devices with a fixed-priority scheduling policy.  The network employs a Carrier Sense Multiple Access with Arbitration on Message Priority (CSMA/AMP) model, which is widely used in industrial Controller Area Network (CAN) bus applications~\cite{Gupta_Networked_2009}. The TrueTime library assumes that higher-level network protocols process messages into packets. 
\par

%\endgroup

%\subsubsection{Validation and Performance Evaluation Methodology}
%\begingroup

%\endgroup

%\endgroup

\subsection{Numerical Simulations}
\par

%\begingroup
\subsubsection{Warning Before Harm Occurs}\label{section:evaluation.validity}
%\begingroup
\begin{table}[!t]
    \renewcommand{\arraystretch}{1.3}
    \caption{Safety constraints considered for the TE case study~\cite{Downs_plant_1993}.}\label{table:safety_constraints}
        \begin{center}
        \begin{tabular}{ c||c||c }
        \hline
        \textbf{Output} & \textbf{Low Limit} & \textbf{High Limit}\\
        \hline \hline
        Reactor Pressure & none & 2895 kPa\\
        \hline
        Reactor Temperature & none & 150 \(^{\circ} C\) \\
        \hline
        Reactor Level & 11.8 \(m^3\) & 21.3 \(m^3\) \\
        \hline
        Product Separator Level & 3.3 \(m^3\) & 9.0 \(m^3\) \\
        \hline
        Stripper Base Level & 3.5 \(m^3\) & 6.6 \(m^3\) \\
        \hline
        \end{tabular}
    \end{center}
\end{table}
 As the objective of our  framework is not to detect attacks, but to generate warnings of potential stealthy attacks before harm occurs, we do not consider previous work on attack detection (such as the work surveyed by Giraldo et al.~\cite{Giraldo_Survey_2018}) a suitable baseline to compare our work against. Furthermore, true/false positive metrics, as traditionally defined in the attack detection literature, are not suitable metrics to evaluate our framework, as it is not meant to replace the existing anomaly detector. Rather, our framework's main utility is in guiding the selection of actions that may reveal a potential intrusion before damage occurs. Therefore, our evaluation consists of demonstrating how our algorithm warns well in advance of potential damage and guides the selection of post-warning actions based on the different thresholds that we designed in Section~\ref{section:instance}. We perform this evaluation using two attack scenarios conforming to the threat model described in Section~\ref{section:system}. This approach is in line with previous work on EWS~\cite{Kalutarage_Sensing_2012,Kalutarage_Towards_2015,Apel_Early_2010,Apel_Towards_2009}. For each scenario, we compare the warnings raised by Algorithm 1 to the attack's ability to cause damage without being detected.
\par
To instantiate the framework, we first approximated the LTI model (\ref{eq:sys.ltimodel}) of the system using a standard system identification technique (\texttt{linmod}) in MATLAB.\@ Table~\ref{table:safety_constraints} shows the safety constraints considered in the present case study, based on the process description provided by Downs and Vogel~\cite{Downs_plant_1993}. We derived an appropriate Kalman filter using MATLAB's built-in \(\texttt{kalman}\) function, and we set an anomaly detection threshold based on a desired false alarm rate of \(\beta = 5\% \). Furthermore, we computed the value of the matrix \(\mathbf{\Pi} \) in Equation (\ref{eq:reachsetx}) for a confidence level \( p=1-\beta =0.95 \).
\par
In a previous work~\cite{Azzam_Efficient_2021}, we demonstrated how to tune the length of the prediction horizon \(K\) in order to maximise the accuracy of the safety checking component of Algorithm 1. The results in~\cite{Azzam_Efficient_2021} show that for the TEP, a \(K = 500\), equivalent to \(\approx 15\) min into the future, guarantees the accuracy of the safety checking. Finally, we set the warnings described in Section~\ref{section:instance} to be such that a low criticality warning is issued if Algorithm 1 returns that damage can happen after \(250\) steps or less, i.e. \(\mathbf{SUSP} \geq 0.004\). A high criticality warning is issued if damage can happen after two steps or less with a \(\mathbf{SUSP} \geq 0.5 \).
\par
\begin{figure*}[!t]
    \centering
    \includegraphics[width = \textwidth,keepaspectratio]{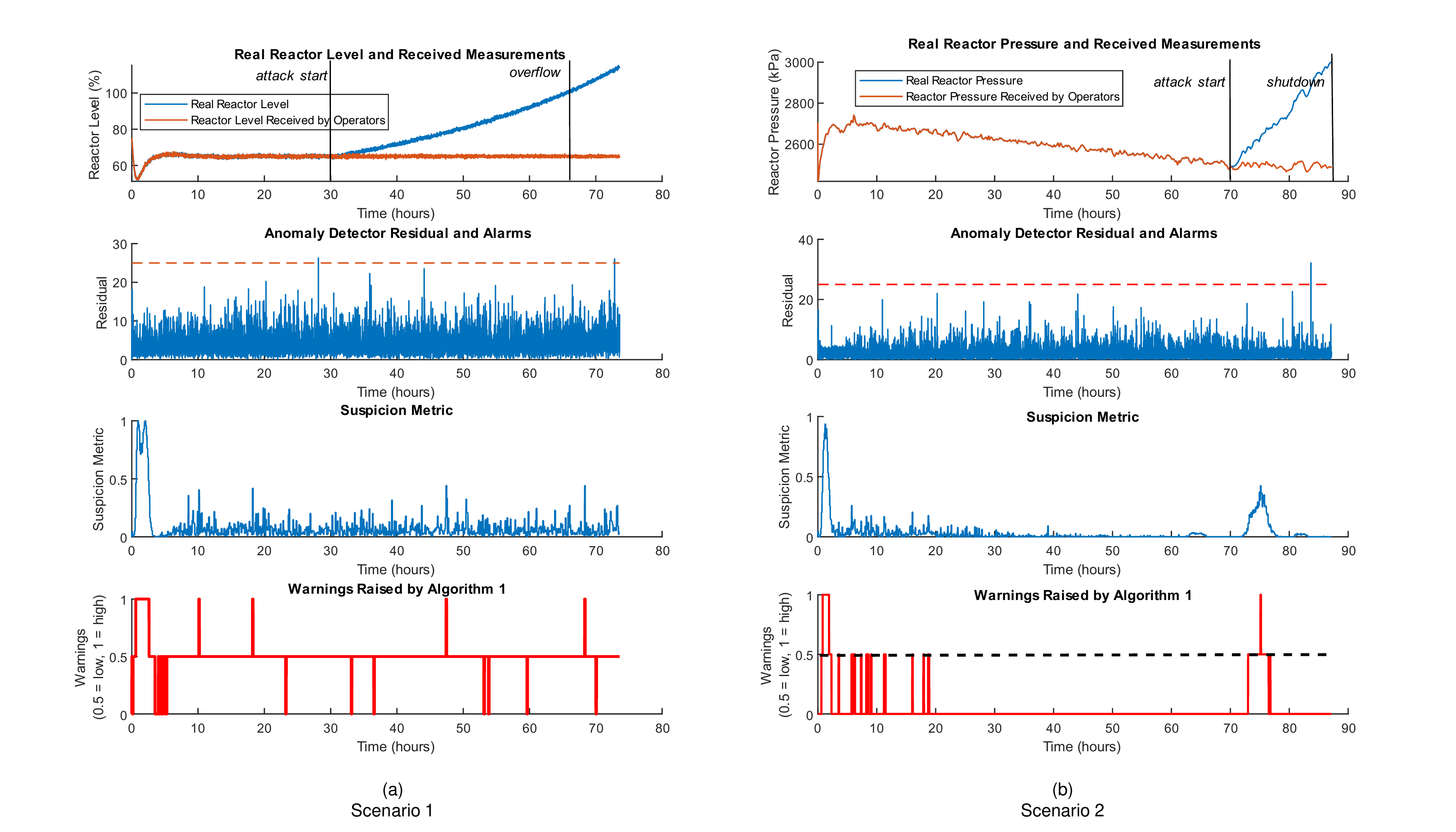}
    \caption{Numerical simulations corresponding to the attack scenarios described in Section~\ref{section:evaluation.validity}. (a) scenario 1; (b) scenario 2.}\label{fig:evalplots}
\end{figure*}

The results obtained for the attack scenarios are shown in Figure~\ref{fig:evalplots}. For each scenario, we plot the value of the reactor level or pressure, residual, suspicion metric, and warning level over time. Our results can be interpreted as follows:
\begin{enumerate}
    \item In the first scenario (Figure~\ref{fig:evalplots}-a), we re-use the same operating conditions and attack described in Section~\ref{section:motivation}. The attack on the reactor starts at \(t=30h\), and the liquids level increases until damage takes place approximately \(37\) hours later. A high-criticality warning is however issued at around \(t=48h\), \(19\) hours before damage takes place. If we relied only on a proximity-based suspicion metric, we would not be able to detect that the reactor was moving to an unsafe state.
    \par
    Before the attack starts, a low-criticality warning level is maintained most of the time. A high-criticality warning is raised after the attack starts, but well before any damage can happen (20 hours). It is worth mentioning that since a low-criticality warning level is maintained for a long period of time, it may imply that actions should be taken all the time which may affect the usability of our approach. However, following this warning level, operators can initially collect more data from the concerned area of the system to confirm or refute the hypothesis that an intrusion is present. If the hypothesis is refuted yet a low-criticality warning level is maintained, then data collection can either stop or take place periodically in order to make sure that no intrusion is present. Collected data can also be stored for a certain amount of time for use in any potential investigation. Moreover, we expect the warnings generated by our framework to be correlated with other alerts (e.g.\ cyber intelligence, insider activity) generated by the EWS. This will provide a more accurate picture of the security situation.  
    \item In the second scenario, we test the ability of Algorithm 1 to provide warning when the plant is attacked during transient operating conditions. We consider a scenario (Figure~\ref{fig:evalplots}-b) where the reactor's pressure is steadily brought to lower levels over a long period of time. The attack on the reactor's pressure starts at \(t=70h\), and excessive pressure starts building up in the reactor until damage takes place approximately \(18\) hours later. A high-criticality warning is however issued at around \(t=75h\), \(13\) hours before damage takes place. Even though the reactor pressure was being lowered throughout the run, Algorithm 1 still identified a state that can be taken to an unsafe operating region through a stealthy attack. If we relied on a purely proximity-based suspicion metric, the reactor's pressure would have appeared to evolve away from an unsafe state.
\end{enumerate}
%We have showed through the previous scenarios that the proposed algorithm is indeed able to warn about stealthy attacks, well before damage takes place. Before illustrating through a use-case scenario the utility of the algorithm in practice, we evaluate its scalability and real-time performance.
%\endgroup

\subsubsection{Scalability and Network Overhead}
\begin{figure}[!t]
    \centering
    \includegraphics[width=0.8\columnwidth,keepaspectratio]{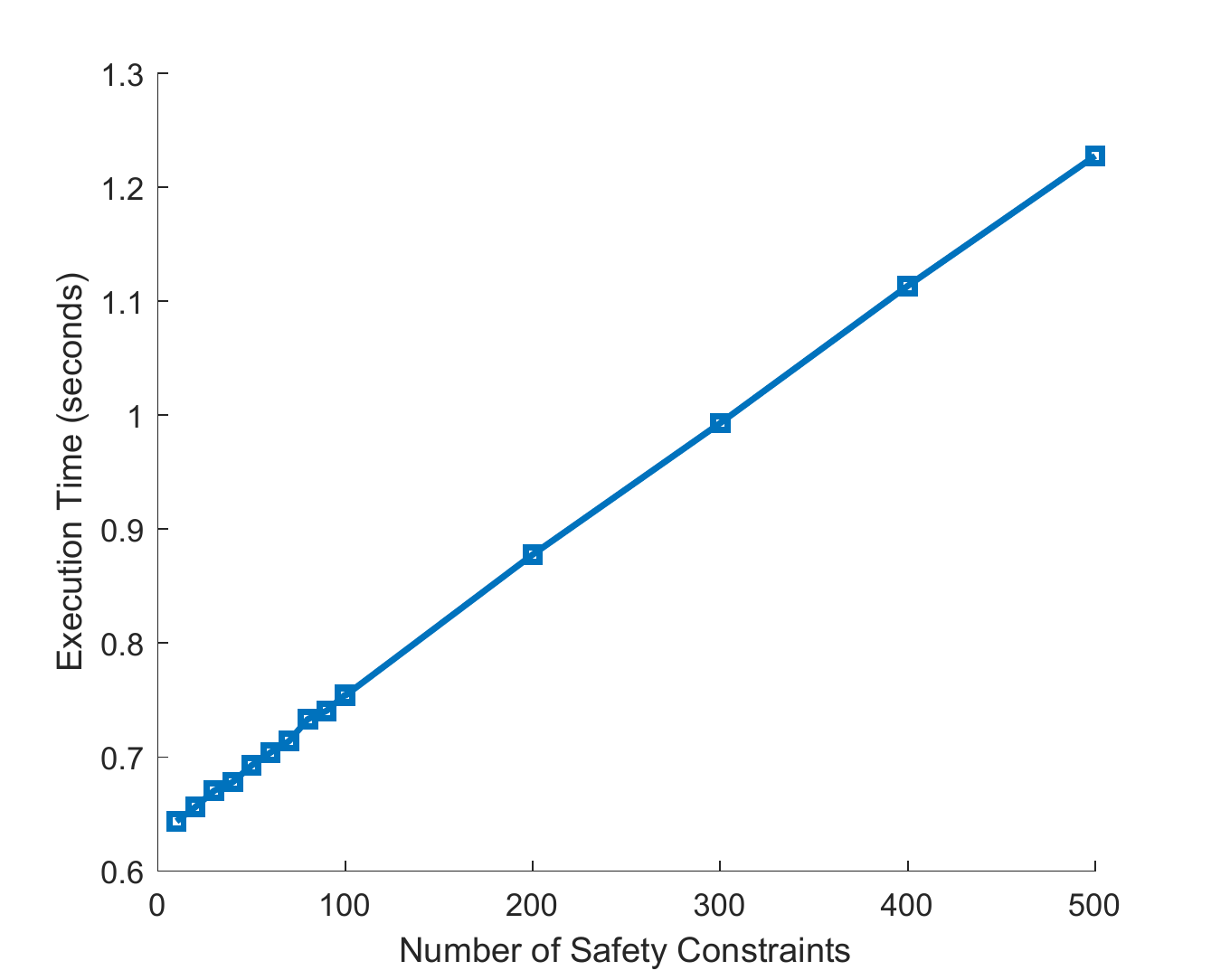}
    \caption{Average execution time of Algorithm 1 vs.\ the number of safety constraints.}\label{fig:results_perfsafetyconst}
\end{figure}
\begin{figure}[!t]
    \centering
    \includegraphics[width = 0.8\columnwidth,keepaspectratio]{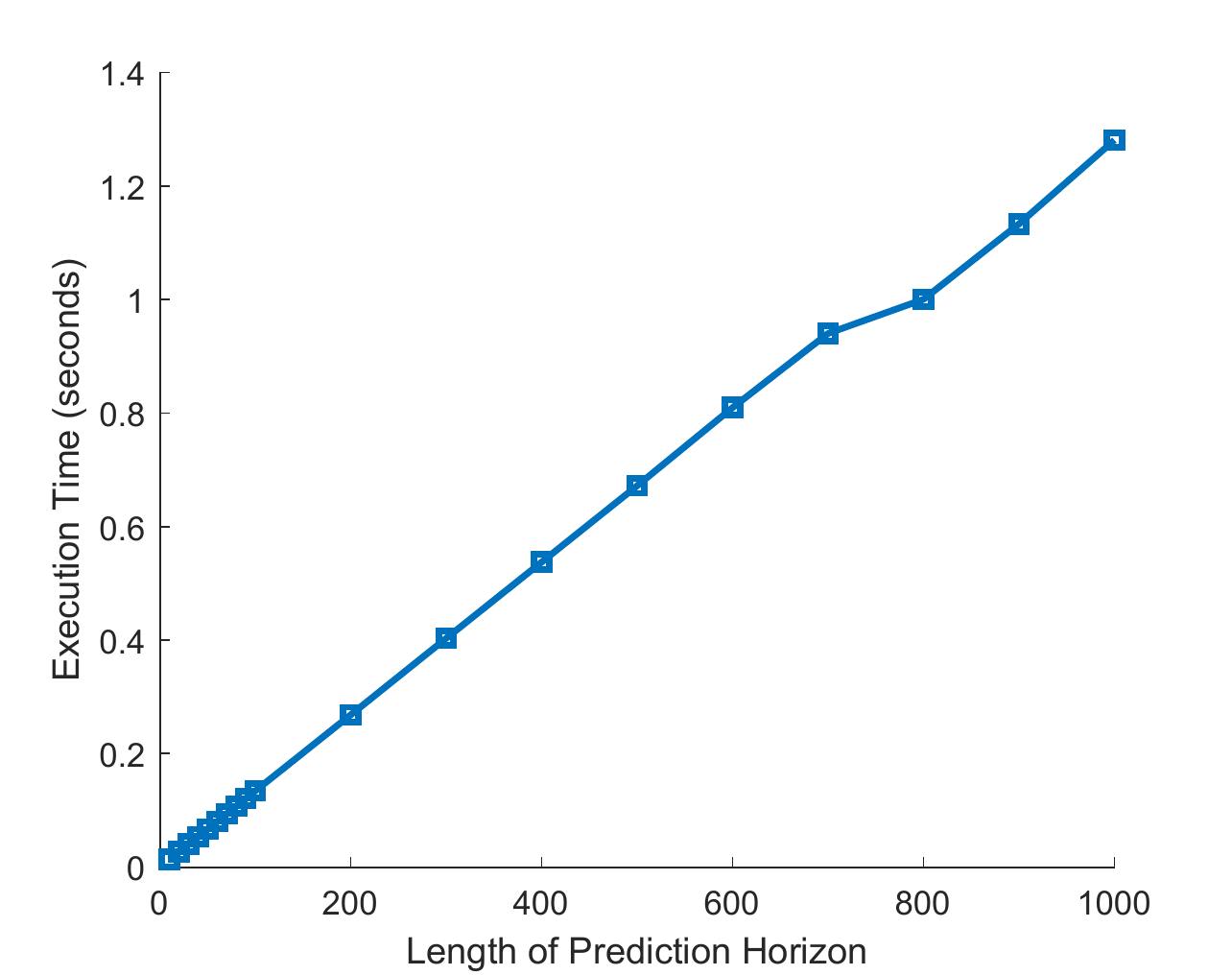}
    \caption{Average execution time of Algorithm 1 vs.\ the length of prediction horizon \(K\).}\label{fig:results_perf}
\end{figure}
%\begingroup
\noindent\textbf{Scalability.} We assessed the scalability of Algorithm 1 with respect to (i) the number of safety constraints, and (ii) the length of the prediction horizon \(K\) set by operators. For both cases, we averaged the execution time over a \(100\)-hour simulation of the networked TEP, equivalent to \(2\times10^5\) executions of the algorithm given the sampling time \(\Delta_t = 5\times 10^{-4}\; \text{hours}\;\approx 1.8 \;\text{seconds}\). In addition, for the purposes of performance testing, we modified Algorithm 1 to simulate the worst-case execution scenario where emptiness checks are performed at every predicted state. To test for scalability against the number of safety constraints, we generated random half-spaces representing potential safety constraints. We also fixed the length of the prediction horizon at \(K = 500\), equivalent to about \(15\) mins into the future. For scalability with the length of the prediction horizon, we used the safety constraints in Table~\ref{table:safety_constraints}. Results are shown in Figures~\ref{fig:results_perfsafetyconst} and~\ref{fig:results_perf}.
\par
The worst-case execution time of the algorithm scales linearly w.r.t. both the number of safety constraints and the length of the prediction horizon. These results prove the ability of the proposed algorithm to scale in safety-critical scenarios, where a larger number of safety constraints are imposed. Furthermore, at \( K = 1000 \) time steps, equivalent to about \(30\) min ahead-of-time prediction, the worst-case execution time \(\approx 1.3\) sec is less than the sampling period, \(1.8\) sec, which  guarantees a satisfactory real-time response. Note that real-time response can be further improved by performing checks only when the estimated physical and controller states of the system (i.e.\ the main real-time inputs to Algorithm 1) are undergoing significant changes, i.e. during transient operation. Finally, the algorithm's worst-case execution time can be improved significantly by considering an implementation in a compiled language such as C rather than an interpreted language like MATLAB.
\par
\begin{table*}[!t]
    \renewcommand{\arraystretch}{1.3}
    \caption{End-to-End transmission time delays in ms between various components of the TEP --- in the presence of the proposed scheme, (without it)}\label{table:perf_timeDelays}
    \begin{center}
        \begin{tabular}{ c||c||c||c||c}
        \hline
        \textbf{Point of Delay Measurement} & \textbf{Average} & \textbf{Standard Deviation} & \textbf{Maximum} & \textbf{Minimum}\\
        \hline \hline
        Sensors to Controllers (in ms) & 6.017 (6.013) & 2.78 (2.77) & 9.65 (9.64) & 1.52 (1.52) \\
        \hline
        Controllers to Actuators (in ms) & 22.90 (19.05) & 9.17 (8.41) & 35.66 (33.91) & 10.19 (1.15)\\
        \hline
        \end{tabular}
    \end{center}
\end{table*}
\noindent\textbf{Network Overhead.} To assess the effect of the proposed algorithm on the performance of the network, we measured end-to-end time delays at two critical locations in the network: (i) between each sensor and its corresponding controller; (ii) between each controller and its corresponding actuator. For each location, we averaged these measurements over a \(100\)-hour simulation. The values in Table~\ref{table:perf_timeDelays} shows the average, standard deviation, maximum, and minimum time delays considering all sensors, controllers, and actuators on the network. The proposed algorithm incurs little additional delays on the network. This result is expected since the installation of the proposed scheme requires that only  sensor values and controller states are uploaded to the supervisory control room area (Figure~\ref{fig:testbed}). These values are already uploaded to perform anomaly detection in the absence of the proposed framework. It is worth noting that these values are usually uploaded also to process historians for various process control and diagnostics-related logging activities. Therefore, the proposed framework is expected to incur little overhead on the network.
%\endgroup
%\endgroup

\subsection{Use Case Scenario}
%\begingroup
\begin{figure}[ht]
    \centering
    \includegraphics[width=\columnwidth,keepaspectratio]{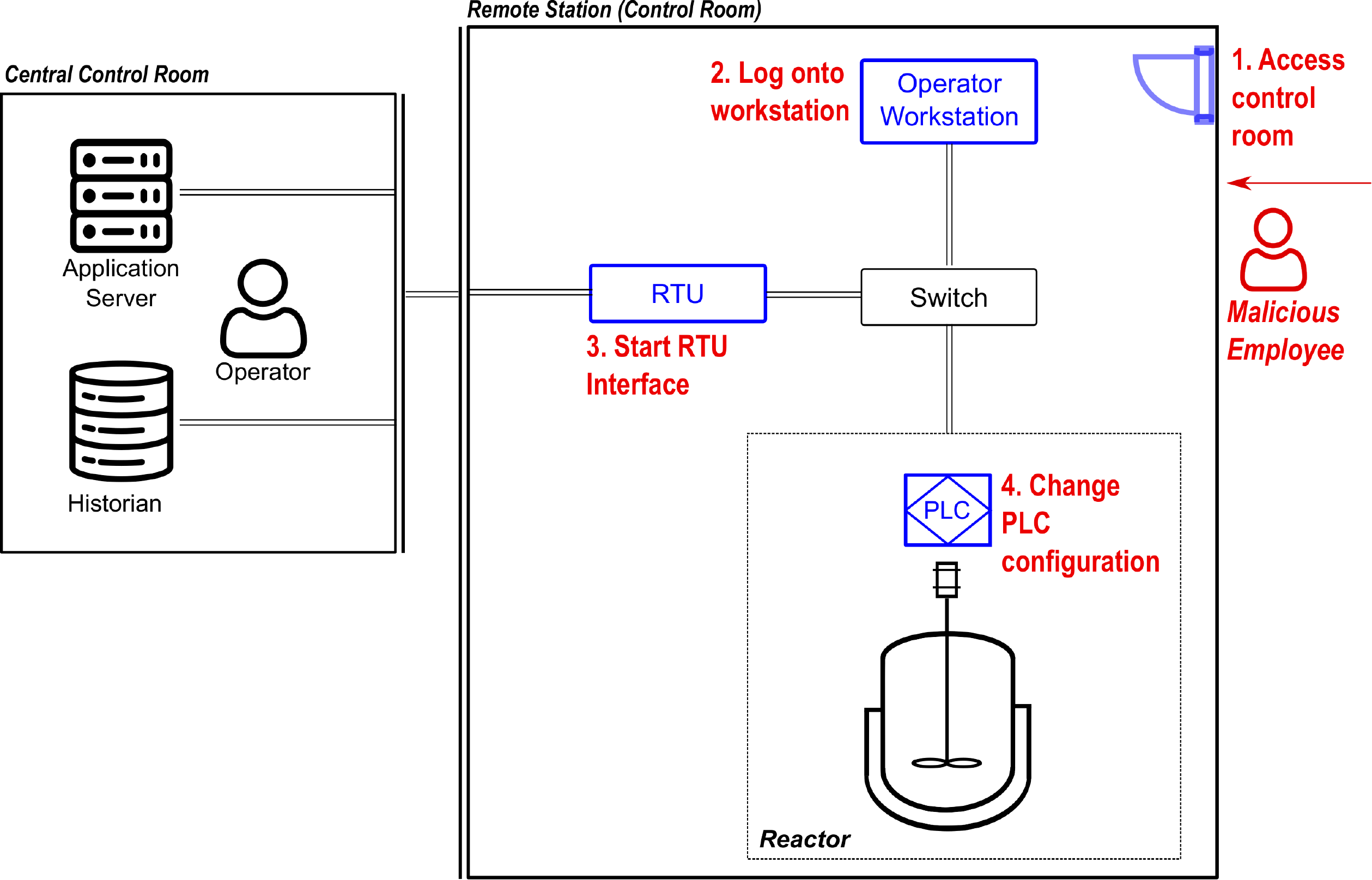}
    \caption{Layout of the system showing the different attack steps (\(y^a(t)=\) fake bias). Sources of potential evidence are highlighted in blue.}\label{fig:system.layout_attackSteps}
\end{figure}
\par
\subsubsection{Layout of the ICS}
For the purposes of this scenario, we restrict our focus on the reactor stage of the TEP. Namely, we assume that the reactor is housed in a remote control station, the layout of which is inspired by the laboratory experiment performed by Sand~\cite{Sand_Incident_2019} and is shown in Figure~\ref{fig:system.layout_attackSteps}. The reactor's temperature controller is installed on a Programmable Logic Controller (PLC), and a Remote Terminal Unit (RTU) is used to interface with the PLC using a speciality software installed on the engineering workstation. The RTU relays control data to the central Supervisory Control And Data Acquisition (SCADA) station, which includes process diagnostics and alarms generated due to possible anomalies.
\par
\subsubsection{Attack Scenario}
Consider the case of a disgruntled employee looking to cause physical damage to the process. We adapt an attack scenario proposed by Sand~\cite{Sand_Incident_2019} which consists of the steps shown in Figure~\ref{fig:system.layout_attackSteps}. The employee uses his access credentials to enter the reactor room and log onto the engineering workstation. Using his knowledge of the physical process, the employee changes the configuration of the PLC such that false data is slowly injected into the temperature sensor following the stealthy attack described in Section~\ref{section:motivation}. Finally, he leaves the reactor room before damage occurs.
\par
\noindent\textbf{Case 1: Low Criticality Warning.}
Following a warning of low criticality, relevant data can be uploaded to the main SCADA server to help profile an alleged attack. Figure~\ref{fig:system.layout_attackSteps} shows potential sources of data in low-level devices:
\begin{enumerate}
    \item \textit{Access control device}: Access control logs would show that the employee was in the remote station as the warning was issued.
    \item \textit{Operator workstation/HMI}: The HMI of the operator workstation stores logs recording log-on events and issued commands which can reveal the sequence of actions performed by the employee. Vendor-specific software~\cite{Eden_Forensic_2016} exists to extract and subsequently upload such logs.
    \item \textit{RTU/PLC}: The current control program executing on the PLC triggers events that can constitute an evidence of the attack performed by the employee. Extracting such information from PLC's without having to power it down is possible, for example, by recording memory variable values using proprietary software such as PLCLogger~\cite{Eden_Forensic_2016}.
\end{enumerate}
By performing this live forensics activity, operators can detect that the employee is initiating a stealthy attack. They may then force the restoration of the PLC configuration and the stored evidence may be used to potentially prosecute the employee in question.
\par
\noindent\textbf{Case 2: High Criticality Warning.}
A high-criticality warning, such as the one shown in Figure~\ref{fig:evalplots}-c would typically be followed by measures to prevent a potential incident, in addition to the data collection activities mentioned previously. In the present scenario, such measures could include engaging a trusted backup PLC to handle the control of the reactor instead of the one with the potentially malicious configuration; or temporarily disabling the RTO module.
%\endgroup

\subsection{Discussion}
%\begingroup
Before concluding this paper, we make some final remarks regarding both the usefulness of the framework and potential practical deployment issues.
\par
\subsubsection{Supporting ICS Forensics}
Recently, an increasing body of work has considered forensics in ICS~\cite{Awad_Tools_2018,AlSharif_Live_2018,Eden_Forensic_2016,Altschaffel_Digital_2019}. In these systems, the low processing power of low-level devices such as sensors and actuators limits the deployment of event-logging tools. In addition, the process' safety criticality limits the degree of interference of forensic tools with the system. It is also often impossible to shutdown an ICS to perform post-mortem forensics, which forces operators to rely mainly on live forensics methods~\cite{Altschaffel_Digital_2019}.
\par
The proposed framework can support live forensics by triggering data collection activities only when a warning about potential damage to equipment is raised. This selective data collection activity, as illustrated in the previous use case scenario, reduces the overhead for the network and the low-processing devices. In addition, measuring grounds for suspicion in real-time reduces the risk of losing evidence about a stealthy attack, in case this attack causes damage to the ICS sensors and/or actuators.
\par
Furthermore, note that in the scenario described previously, the attacker did not need to break into the network or take advantage of a software vulnerability. An EWS relying only on indicators based on network events may not be able to warn well in advance of such an attack. Our framework can complement an EWS that monitors insider activity (e.g.~\cite{Chivers_Accumulating_2009}) by warning when such activity targets the physical components and may cause damage.

\subsubsection{Potential Limitations}
First, the configuration of our framework may require some manual effort. However, we remark that approximate mathematical models of the system, its anomaly detector, and unsafe operation are standard in control engineering. Second, to alleviate the computational cost incurred by reachability analysis, we computed approximate symbolic reachable sets offline and instantiated them at runtime --- an approach inspired by simplex control architectures~\cite{Chen_Model_2017}. We realise nonetheless that the reachability tools we used in the present case study are specific to LTI systems. To increase generalizability of our results, in future work we will explore adoption of a different set of tools  to instantiate our framework to different types of systems. Finally, to reduce the risk of a potential adversary subverting the EWS, one possible solution is to implement it on a Shadow Security Unit (SSU) as proposed by Graveto et al.~\cite{Graveto_stealth_2019}. Such devices are computers designed specifically to remain hidden from potential offenders. Encryption mechanisms can also be added as an extra layer of security, to favour more secure communications at the cost of potentially reduced performance.
%\endgroup
%%% Local Variables:
%%% mode: latex
%%% TeX-master: "main"
%%% End:

%% file: section8.Conclusion.tex
\section{Conclusion}\label{section:concl}
%\begingroup
In this paper, we considered the problem of stealthy attacks on
safety-critical ICS.\ We proposed a framework which can be used as part of an EWS to raise early warnings
based on grounds for suspicion representing ``physics-based'' preliminary indicators of a stealthy attack. We defined two grounds for suspicion based on
the physical dynamics of a system: (i) feasibility of a stealthy
attack and (ii) proximity of the system to unsafe operating
regions. To monitor the grounds for suspicion  in
real-time, we  proposed a suspicion metric based on a mathematical
model of the system. We also provided soundness principles to ensure
that the metric is consistent with the measured grounds.
%
% and a suspicion
% metric which maps these cues in real-time to a value that provides a
% measure of the likelihood of success of a stealthy attack. We also
% proposed soundness principles to ensure a meaningful instantiation of
% our metric for a particular system and threat model.
% \par 
%\par
% We propose a framework to generate early warnings in ICS based on prelim- inary indicators of a stealthy attack, referred to as grounds for suspicion. The success of a stealthy attack depends on the laws of physics underlying the be- haviour of the ICS and the anomaly detector. Thus, we define for suspicion based on the physical state of the system. i) Feasibility of a stealthy attack indicates whether the ICS can be taken to an unsafe operating region, while avoiding detection by the IDS. ii) Proximity of a stealthy attacks repre- sents the vicinity of the system to the unsafe operating region. Intuitively, a higher proximity can indicate an increased chance of an attacker to stealthily damage the system. 
To illustrate our framework, we considered the case of a
safety-critical chemical reactor system faced with a stealthy attack
on its sensors. We adapted reachability tools from the literature,
namely ellipsoidal calculus, to evaluate the suspicion metric
efficiently in real-time. We also illustrated with a use case that
our framework can support live forensics activities, by triggering early
evidence collection and preserving potential evidence about a stealthy
attack.
\par 
Going forward, we aim to apply our framework to
different systems and attacks. We will consider other  existing
threat models and  modelling frameworks for the physical process. We will also
implement a prototype tool to reduce the human effort required to
instantiate our framework to a specific system. Finally, we will
further investigate the applicability of our work to ICS
forensics.
%\endgroup
%%% Local Variables:
%%% mode: latex
%%% TeX-master: "main"
%%% End:

%% file: main.bbl
\begin{thebibliography}{10}
\providecommand{\url}[1]{#1}
\csname url@samestyle\endcsname
\providecommand{\newblock}{\relax}
\providecommand{\bibinfo}[2]{#2}
\providecommand{\BIBentrySTDinterwordspacing}{\spaceskip=0pt\relax}
\providecommand{\BIBentryALTinterwordstretchfactor}{4}
\providecommand{\BIBentryALTinterwordspacing}{\spaceskip=\fontdimen2\font plus
\BIBentryALTinterwordstretchfactor\fontdimen3\font minus
  \fontdimen4\font\relax}
\providecommand{\BIBforeignlanguage}[2]{{%
\expandafter\ifx\csname l@#1\endcsname\relax
\typeout{** WARNING: IEEEtran.bst: No hyphenation pattern has been}%
\typeout{** loaded for the language `#1'. Using the pattern for}%
\typeout{** the default language instead.}%
\else
\language=\csname l@#1\endcsname
\fi
#2}}
\providecommand{\BIBdecl}{\relax}
\BIBdecl

\bibitem{Alur_Principles_2015}
R.~Alur, \emph{Principles of cyber-physical systems}.\hskip 1em plus 0.5em
  minus 0.4em\relax MIT Press, 2015.

\bibitem{Lee_SANS_2014}
R.~Lee, M.~Assante, and T.~Conway, \emph{SANS ICS Defense Use Case (DUC) Dec 30
  2014: ICS CP/PE case study paper-German Steel Mill Cyber Attack}.\hskip 1em
  plus 0.5em minus 0.4em\relax SANS ICS, 2014.

\bibitem{Lee_Analysis_2016}
R.~M. Lee, M.~J. Assante, and T.~Conway, ``Analysis of the cyber attack on the
  ukrainian power grid: Defense use case,'' \emph{SANS ICS}, 2016.

\bibitem{Giraldo_Survey_2018}
J.~Giraldo, D.~Urbina, A.~Cardenas, J.~Valente, M.~Faisal, J.~Ruths, N.~O.
  Tippenhauer, H.~Sandberg, and R.~Candell, ``A survey of physics-based attack
  detection in cyber-physical systems,'' \emph{ACM Computing Surveys (CSUR)},
  vol.~51, no.~4, p.~76, 2018.

\bibitem{Bai_Security_2015}
C.-Z. Bai, F.~Pasqualetti, and V.~Gupta, ``Security in stochastic control
  systems: Fundamental limitations and performance bounds,'' in \emph{American
  Control Conference (ACC), 2015}.\hskip 1em plus 0.5em minus 0.4em\relax IEEE,
  2015, Conference Proceedings, pp. 195--200.

\bibitem{Pasqualetti_Attack_2013}
F.~Pasqualetti, F.~Dörfler, and F.~Bullo, ``Attack detection and
  identification in cyber-physical systems,'' \emph{IEEE Transactions on
  Automatic Control}, vol.~58, no.~11, pp. 2715--2729, 2013.

\bibitem{Apel_Towards_2009}
M.~Apel, J.~Biskup, U.~Flegel, and M.~Meier, ``Towards early warning
  systems–challenges, technologies and architecture,'' in \emph{International
  Workshop on Critical Information Infrastructures Security}.\hskip 1em plus
  0.5em minus 0.4em\relax Springer, 2009, Conference Proceedings, pp. 151--164.

\bibitem{Kalutarage_Early_2015}
H.~Kalutarage, S.~Shaikh, B.-S. Lee, C.~Lee, and Y.~C. Kiat, ``Early warning
  systems for cyber defence,'' in \emph{International Workshop on Open Problems
  in Network Security}.\hskip 1em plus 0.5em minus 0.4em\relax Springer, 2015,
  Conference Proceedings, pp. 29--42.

\bibitem{Ramaki_Survey_2016}
A.~A. Ramaki and R.~E. Atani, ``A survey of it early warning systems:
  architectures, challenges, and solutions,'' \emph{Security and Communication
  Networks}, vol.~9, no.~17, pp. 4751--4776, 2016.

\bibitem{Murguia_reachable_2018}
C.~Murguia and J.~Ruths, ``On reachable sets of hidden cps sensor attacks,'' in
  \emph{2018 Annual American Control Conference (ACC)}.\hskip 1em plus 0.5em
  minus 0.4em\relax IEEE, 2018, Conference Proceedings, pp. 178--184.

\bibitem{Chen_Model_2017}
X.~Chen and S.~Sankaranarayanan, ``Model predictive real-time monitoring of
  linear systems,'' in \emph{2017 IEEE Real-Time Systems Symposium
  (RTSS)}.\hskip 1em plus 0.5em minus 0.4em\relax IEEE, 2017, Conference
  Proceedings, pp. 297--306.

\bibitem{kurzhanski_ellipsoidal_2000}
A.~B. Kurzhanski and P.~Varaiya, ``Ellipsoidal techniques for reachability
  analysis,'' in \emph{International Workshop on Hybrid Systems: Computation
  and Control}.\hskip 1em plus 0.5em minus 0.4em\relax Springer, 2000, pp.
  202--214.

\bibitem{kurzhanskiy_ellipsoidal_2006}
A.~A. Kurzhanskiy and P.~Varaiya, ``Ellipsoidal toolbox (et),'' in
  \emph{Proceedings of the 45th IEEE Conference on Decision and Control}.\hskip
  1em plus 0.5em minus 0.4em\relax IEEE, 2006, pp. 1498--1503.

\bibitem{boyd_convex_2004}
S.~Boyd, S.~P. Boyd, and L.~Vandenberghe, \emph{Convex optimization}.\hskip 1em
  plus 0.5em minus 0.4em\relax Cambridge university press, 2004.

\bibitem{Azzam_Efficient_2021}
M.~Azzam, L.~Pasquale, G.~Provan, and B.~Nuseibeh, ``Efficient predictive
  monitoring of linear time-invariant systems under stealthy attacks,''
  \emph{arXiv preprint arXiv 2106.02378}, 2021.

\bibitem{Genge_connection_2014}
B.~Genge, D.~A. Rusu, and P.~Haller, ``A connection pattern-based approach to
  detect network traffic anomalies in critical infrastructures,'' in
  \emph{Proceedings of the Seventh European Workshop on System Security}.\hskip
  1em plus 0.5em minus 0.4em\relax ACM, 2014, Conference Proceedings, p.~1.

\bibitem{Cheminod_Detection_2017}
M.~Cheminod, L.~Durante, L.~Seno, and A.~Valenzano, ``Detection of attacks
  based on known vulnerabilities in industrial networked systems,''
  \emph{journal of information security and applications}, vol.~34, pp.
  153--165, 2017.

\bibitem{Sayegh_SCADA_2014}
N.~Sayegh, I.~H. Elhajj, A.~Kayssi, and A.~Chehab, ``Scada intrusion detection
  system based on temporal behavior of frequent patterns,'' in \emph{MELECON
  2014-2014 17th IEEE Mediterranean Electrotechnical Conference}.\hskip 1em
  plus 0.5em minus 0.4em\relax IEEE, 2014, Conference Proceedings, pp.
  432--438.

\bibitem{Mercaldo_Real_2019}
F.~Mercaldo, F.~Martinelli, and A.~Santone, ``Real-time scada attack detection
  by means of formal methods,'' in \emph{2019 IEEE 28th International
  Conference on Enabling Technologies: Infrastructure for Collaborative
  Enterprises (WETICE)}.\hskip 1em plus 0.5em minus 0.4em\relax IEEE, 2019,
  Conference Proceedings, pp. 231--236.

\bibitem{Teixeira_Revealing_2012}
A.~Teixeira, I.~Shames, H.~Sandberg, and K.~H. Johansson, ``Revealing stealthy
  attacks in control systems,'' in \emph{50th Annual Allerton Conference on
  Communication, Control, and Computing, Allerton, IL, USA, October 01-05,
  2012}.\hskip 1em plus 0.5em minus 0.4em\relax IEEE conference proceedings,
  2012, Conference Proceedings, pp. 1806--1813.

\bibitem{Griffioen_Tutorial_2019}
P.~Griffioen, S.~Weerakkody, B.~Sinopoli, O.~Ozel, and Y.~Mo, ``A tutorial on
  detecting security attacks on cyber-physical systems,'' in \emph{2019 18th
  European Control Conference (ECC)}.\hskip 1em plus 0.5em minus 0.4em\relax
  IEEE, 2019, Conference Proceedings, pp. 979--984.

\bibitem{Weerakkody_Active_2017}
S.~Weerakkody, O.~Ozel, P.~Griffioen, and B.~Sinopoli, ``Active detection for
  exposing intelligent attacks in control systems,'' in \emph{2017 IEEE
  Conference on Control Technology and Applications (CCTA)}.\hskip 1em plus
  0.5em minus 0.4em\relax IEEE, 2017, Conference Proceedings, pp. 1306--1312.

\bibitem{Hoehn_Detection_2016}
A.~Hoehn and P.~Zhang, ``Detection of covert attacks and zero dynamics attacks
  in cyber-physical systems,'' in \emph{2016 American Control Conference
  (ACC)}.\hskip 1em plus 0.5em minus 0.4em\relax IEEE, 2016, Conference
  Proceedings, pp. 302--307.

\bibitem{Apel_Early_2010}
M.~Apel, J.~Biskup, U.~Flegel, and M.~Meier, \emph{Early Warning System on a
  National Level: Project AMSEL}.\hskip 1em plus 0.5em minus 0.4em\relax
  Universitätsbibliothek Dortmund, 2010.

\bibitem{Kalutarage_How_2013}
H.~K. Kalutarage, S.~A. Shaikh, Q.~Zhou, and A.~E. James, ``How do we
  effectively monitor for slow suspicious activities?'' in \emph{ESSoS Doctoral
  Symposium 2013}.\hskip 1em plus 0.5em minus 0.4em\relax Citeseer, 2013,
  Conference Proceedings, p.~36.

\bibitem{Kalutarage_Towards_2015}
H.~K. Kalutarage, C.~Lee, S.~A. Shaikh, and F.~L.~B. Sung, ``Towards an early
  warning system for network attacks using bayesian inference,'' in \emph{2015
  IEEE 2nd International Conference on Cyber Security and Cloud
  Computing}.\hskip 1em plus 0.5em minus 0.4em\relax IEEE, 2015, Conference
  Proceedings, pp. 399--404.

\bibitem{Abbaszadeh_Forecasting_2018}
M.~Abbaszadeh, L.~K. Mestha, and W.~Yan, ``Forecasting and early warning for
  adversarial targeting in industrial control systems,'' in \emph{2018 IEEE
  Conference on Decision and Control (CDC)}, 2018, Conference Proceedings, pp.
  7200--7205.

\bibitem{Etigowni_Crystal_2018}
S.~Etigowni, S.~Hossain-McKenzie, M.~Kazerooni, K.~Davis, and S.~Zonouz,
  ``Crystal (ball) i look at physics and predict control flow!
  just-ahead-of-time controller recovery,'' in \emph{Proceedings of the 34th
  Annual Computer Security Applications Conference}, 2018, Conference
  Proceedings, pp. 553--565.

\bibitem{Carcano_multidimensional_2011}
A.~Carcano, A.~Coletta, M.~Guglielmi, M.~Masera, I.~N. Fovino, and
  A.~Trombetta, ``A multidimensional critical state analysis for detecting
  intrusions in scada systems,'' \emph{IEEE Transactions on Industrial
  Informatics}, vol.~7, no.~2, pp. 179--186, 2011.

\bibitem{Coletta_Predictive_2018}
A.~Coletta, ``Predictive detection of known security criticalities in cyber
  physical systems with unobservable variables,'' in \emph{11th international
  conference on security and its applications (cnsa)}, 2018, Conference
  Proceedings, pp. 61--77.

\bibitem{Castellanos_modular_2019}
J.~H. Castellanos and J.~Zhou, ``A modular hybrid learning approach for
  black-box security testing of cps,'' in \emph{International Conference on
  Applied Cryptography and Network Security}.\hskip 1em plus 0.5em minus
  0.4em\relax Springer, 2019, Conference Proceedings, pp. 196--216.

\bibitem{Bradford_Towards_2004}
P.~G. Bradford, M.~Brown, J.~Perdue, and B.~Self, ``Towards proactive
  computer-system forensics,'' in \emph{International Conference on Information
  Technology: Coding and Computing, 2004. Proceedings. ITCC 2004.},
  vol.~2.\hskip 1em plus 0.5em minus 0.4em\relax IEEE, 2004, Conference
  Proceedings, pp. 648--652.

\bibitem{Chivers_Accumulating_2009}
H.~Chivers, P.~Nobles, S.~A. Shaikh, J.~A. Clark, and H.~Chen, ``Accumulating
  evidence of insider attacks,'' in \emph{Proceedings of the 1st International
  Workshop on Managing Insider Security Threats (MIST-2009)}.\hskip 1em plus
  0.5em minus 0.4em\relax CEUR, 2009, Conference Proceedings, pp. 34--50.

\bibitem{Kalutarage_Sensing_2012}
H.~K. Kalutarage, S.~A. Shaikh, Q.~Zhou, and A.~E. James, ``Sensing for
  suspicion at scale: A bayesian approach for cyber conflict attribution and
  reasoning,'' in \emph{2012 4th International Conference on Cyber Conflict
  (CYCON 2012)}.\hskip 1em plus 0.5em minus 0.4em\relax IEEE, 2012, Conference
  Proceedings, pp. 1--19.

\bibitem{Murguia_Security_2018}
C.~Murguia, I.~Shames, J.~Ruths, and D.~Nesic, ``Security metrics of networked
  control systems under sensor attacks (extended preprint),'' \emph{arXiv
  preprint arXiv:1809.01808}, 2018.

\bibitem{Kwon_Reachability_2018}
C.~Kwon and I.~Hwang, ``Reachability analysis for safety assurance of
  cyber-physical systems against cyber attacks,'' \emph{IEEE Transactions on
  Automatic Control}, vol.~63, no.~7, pp. 2272--2279, 2018.

\bibitem{Sequeira_Real_2002}
S.~E. Sequeira, M.~Graells, and L.~Puigjaner, ``Real-time evolution for on-line
  optimization of continuous processes,'' \emph{Industrial and engineering
  chemistry research}, vol.~41, no.~7, pp. 1815--1825, 2002.

\bibitem{Milosevic_Quantifying_2018}
J.~Milošević, D.~Umsonst, H.~Sandberg, and K.~H. Johansson, ``Quantifying the
  impact of cyber-attack strategies for control systems equipped with an
  anomaly detector,'' in \emph{2018 European Control Conference (ECC)}.\hskip
  1em plus 0.5em minus 0.4em\relax IEEE, 2018, Conference Proceedings, pp.
  331--337.

\bibitem{Urbina_Limiting_2016}
D.~I. Urbina, J.~A. Giraldo, A.~A. Cardenas, N.~O. Tippenhauer, J.~Valente,
  M.~Faisal, J.~Ruths, R.~Candell, and H.~Sandberg, ``Limiting the impact of
  stealthy attacks on industrial control systems,'' in \emph{Proceedings of the
  2016 ACM SIGSAC Conference on Computer and Communications Security}.\hskip
  1em plus 0.5em minus 0.4em\relax ACM, 2016, Conference Proceedings, pp.
  1092--1105.

\bibitem{Aoudi_Truth_2018}
W.~Aoudi, M.~Iturbe, and M.~Almgren, ``Truth will out: Departure-based
  process-level detection of stealthy attacks on control systems,'' in
  \emph{Proceedings of the 2018 ACM SIGSAC Conference on Computer and
  Communications Security}.\hskip 1em plus 0.5em minus 0.4em\relax ACM, 2018,
  Conference Proceedings, pp. 817--831.

\bibitem{Erba_No_2020}
A.~Erba and N.~O. Tippenhauer, ``No need to know physics: Resilience of
  process-based model-free anomaly detection for industrial control systems,''
  \emph{arXiv preprint arXiv:2012.03586}, 2020.

\bibitem{Huang_Understanding_2009}
Y.-L. Huang, A.~A. Cárdenas, S.~Amin, Z.-S. Lin, H.-Y. Tsai, and S.~Sastry,
  ``Understanding the physical and economic consequences of attacks on control
  systems,'' \emph{International Journal of Critical Infrastructure
  Protection}, vol.~2, no.~3, pp. 73--83, 2009.

\bibitem{Teixeira_Secure_2015a}
A.~Teixeira, I.~Shames, H.~Sandberg, and K.~H. Johansson, ``A secure control
  framework for resource-limited adversaries,'' \emph{Automatica}, vol.~51, pp.
  135--148, 2015.

\bibitem{Polisetty_Error_2019}
\BIBentryALTinterwordspacing
V.~G. Polisetty, S.~K. Varanasi, and P.~Jampana, ``Error bounds for
  identification of a class of continuous lti systems,''
  \emph{IFAC-PapersOnLine}, vol.~52, no.~1, pp. 418 -- 423, 2019, 12th IFAC
  Symposium on Dynamics and Control of Process Systems, including Biosystems
  DYCOPS 2019. [Online]. Available:
  \url{http://www.sciencedirect.com/science/article/pii/S2405896319301843}
\BIBentrySTDinterwordspacing

\bibitem{Hashemi_comparison_2018}
N.~Hashemi, C.~Murguia, and J.~Ruths, ``A comparison of stealthy sensor attacks
  on control systems,'' in \emph{2018 Annual American Control Conference
  (ACC)}.\hskip 1em plus 0.5em minus 0.4em\relax IEEE, 2018, Conference
  Proceedings, pp. 973--979.

\bibitem{Downs_plant_1993}
J.~J. Downs and E.~F. Vogel, ``A plant-wide industrial process control
  problem,'' \emph{Computers and chemical engineering}, vol.~17, no.~3, pp.
  245--255, 1993.

\bibitem{Ricker_Model_1993}
N.~L. Ricker, ``Model predictive control of a continuous, nonlinear, two-phase
  reactor,'' \emph{Journal of Process Control}, vol.~3, no.~2, pp. 109--123,
  1993.

\bibitem{Ricker_decentralized_1996}
------, ``Decentralized control of the tennessee eastman challenge process,''
  \emph{Journal of Process Control}, vol.~6, no.~4, pp. 205--221, 1996.

\bibitem{Geng_survey_2019}
Y.~Geng, Y.~Wang, W.~Liu, Q.~Wei, K.~Liu, and H.~Wu, ``A survey of industrial
  control system testbeds,'' in \emph{IOP Conference Series: Materials Science
  and Engineering}, vol. 569.\hskip 1em plus 0.5em minus 0.4em\relax IOP
  Publishing, 2019, Conference Proceedings, p. 042030.

\bibitem{Genge_Physical_2014}
B.~Genge and C.~Siaterlis, ``Physical process resilience-aware network design
  for scada systems,'' \emph{Computers and Electrical Engineering}, vol.~40,
  no.~1, pp. 142--157, 2014.

\bibitem{Krotofil_Resilience_2013}
M.~Krotofil and A.~A. Cárdenas, ``Resilience of process control systems to
  cyber-physical attacks,'' in \emph{Nordic Conference on Secure IT
  Systems}.\hskip 1em plus 0.5em minus 0.4em\relax Springer, 2013, Conference
  Proceedings, pp. 166--182.

\bibitem{McEvoy_plant_2011}
T.~McEvoy and S.~Wolthusen, ``A plant-wide industrial process control security
  problem,'' in \emph{International Conference on Critical Infrastructure
  Protection}.\hskip 1em plus 0.5em minus 0.4em\relax Springer, 2011,
  Conference Proceedings, pp. 47--56.

\bibitem{Cardenas_Attacks_2011}
A.~A. Cárdenas, S.~Amin, Z.-S. Lin, Y.-L. Huang, C.-Y. Huang, and S.~Sastry,
  ``Attacks against process control systems: risk assessment, detection, and
  response,'' in \emph{Proceedings of the 6th ACM symposium on information,
  computer and communications security}.\hskip 1em plus 0.5em minus 0.4em\relax
  ACM, 2011, Conference Proceedings, pp. 355--366.

\bibitem{Bathelt_2015_Revision}
A.~Bathelt, N.~L. Ricker, and M.~Jelali, ``Revision of the tennessee eastman
  process model,'' \emph{IFAC-PapersOnLine}, vol.~48, no.~8, pp. 309--314,
  2015.

\bibitem{Cervin_Does_2003}
A.~Cervin, D.~Henriksson, B.~Lincoln, J.~Eker, and K.-E. Arzen, ``How does
  control timing affect performance? analysis and simulation of timing using
  jitterbug and truetime,'' \emph{IEEE control systems magazine}, vol.~23,
  no.~3, pp. 16--30, 2003.

\bibitem{Kalaivani_Earliest_2019}
C.~Kalaivani and N.~Kalaiarasi, ``Earliest deadline first scheduling technique
  for different networks in network control system,'' \emph{Neural Computing
  and Applications}, vol.~31, no.~1, pp. 223--232, 2019.

\bibitem{Brahimi_Comparison_2006}
B.~Brahimi, E.~Rondeau, and C.~Aubrun, ``Comparison between networked control
  system behaviour based on can and switched ethernet networks,'' \emph{arXiv
  preprint cs/0611149}, 2006.

\bibitem{Farooqui_Cyber_2014}
A.~A. Farooqui, S.~S.~H. Zaidi, A.~Y. Memon, and S.~Qazi, ``Cyber security
  backdrop: A scada testbed,'' in \emph{2014 IEEE Computers, Communications and
  IT Applications Conference}.\hskip 1em plus 0.5em minus 0.4em\relax IEEE,
  2014, Conference Proceedings, pp. 98--103.

\bibitem{Gupta_Networked_2009}
R.~A. Gupta and M.-Y. Chow, ``Networked control system: Overview and research
  trends,'' \emph{IEEE transactions on industrial electronics}, vol.~57, no.~7,
  pp. 2527--2535, 2009.

\bibitem{Sand_Incident_2019}
K.~A. Sand, ``Incident handling, forensics sensors and information sources in
  industrial control systems,'' Thesis, NTNU, 2019.

\bibitem{Eden_Forensic_2016}
P.~Eden, A.~Blyth, P.~Burnap, Y.~Cherdantseva, K.~Jones, H.~Soulsby, and
  K.~Stoddart, ``Forensic readiness for scada/ics incident response,'' in
  \emph{Proceedings of the 4th International Symposium for ICS and SCADA Cyber
  Security Research 2016}.\hskip 1em plus 0.5em minus 0.4em\relax BCS Learning
  and Development Ltd., 2016, Conference Proceedings, pp. 1--9.

\bibitem{Awad_Tools_2018}
R.~A. Awad, S.~Beztchi, J.~M. Smith, B.~Lyles, and S.~Prowell, ``Tools,
  techniques, and methodologies: A survey of digital forensics for scada
  systems,'' in \emph{Proceedings of the 4th Annual Industrial Control System
  Security Workshop}.\hskip 1em plus 0.5em minus 0.4em\relax ACM, 2018,
  Conference Proceedings, pp. 1--8.

\bibitem{AlSharif_Live_2018}
Z.~A. Al-Sharif, M.~I. Al-Saleh, L.~M. Alawneh, Y.~I. Jararweh, and B.~Gupta,
  ``Live forensics of software attacks on cyber–physical systems,''
  \emph{Future Generation Computer Systems}, 2018.

\bibitem{Altschaffel_Digital_2019}
R.~Altschaffel, M.~Hildebrandt, S.~Kiltz, and J.~Dittmann, ``Digital forensics
  in industrial control systems,'' in \emph{International Conference on
  Computer Safety, Reliability, and Security}.\hskip 1em plus 0.5em minus
  0.4em\relax Springer, 2019, Conference Proceedings, pp. 128--136.

\bibitem{Graveto_stealth_2019}
V.~Graveto, L.~Rosa, T.~Cruz, and P.~Simões, ``A stealth monitoring mechanism
  for cyber-physical systems,'' \emph{International Journal of Critical
  Infrastructure Protection}, vol.~24, pp. 126--143, 2019.

\end{thebibliography}
